\documentclass[aps,prl,10pt,twocolumn,groupedaddress,showkeys,longbibliography]{revtex4-1}
\usepackage{amsmath}
\usepackage{graphicx}
\usepackage{color}

\begin{document}

\title{Resonant field emission from noble-metal/graphene heterostructures}

\author{Maxim Trushin}
\email{mxt@nus.edu.sg}
\affiliation{Institute for Functional Intelligent Materials, National University of Singapore, 4 Science Drive 2, Singapore 117544, Singapore,}
\affiliation{Department of Materials Science and Engineering, National University of Singapore, 9 Engineering Drive 1, Singapore 117575, Singapore}


\begin{abstract}
Field emission from metals underpinned early vacuum-tube technology, 
and recent nanoscale engineering made field-emission devices compatible with modern silicon platforms.
However, the limited tunability of electron transport in metals has restricted their applicability.
Here, we show that noble metals coated with graphene exhibit clean non-monotonic $I-V$ characteristics arising from resonant tunneling through graphene's electronic states, enabled by graphene's atomic thinness and weak electronic hybridization with noble metals.
Our approach combines ab-initio interface parameters with exact solutions of the Schr\"odinger equation for electron transmission across the interface.
We analyze two experimentally relevant geometries: a vertical configuration with a flat suspended emitter and a coplanar configuration with sharp electrodes allowing for strong field enhancement and gating.
These results establish a practical route to tunable electron transport in metal heterostructures, positioning them as competitive components for air-channel field-emission nanoelectronics.
\end{abstract}

\maketitle

Field emission has been studied for more than a century \cite{thomson1897xl}, with Fowler-Nordheim (FN) theory describing electron tunneling from metallic conduction-band states under strong electric fields \cite{fowler1928electron}.  When vacuum channels are downscaled to tens of nanometers  \cite{han2017nanoscale,han2019nanoscale,nirantar2018metal,turchetti2022electron,heo2023vacuum}, and even below 10 nm \cite{wang2024enhancing},
measured currents generally follow FN scaling, with platinum contacts as a notable exception \cite{nirantar2018metal}.
Resonant tunneling, initially introduced by placing a single atom at an emitter tip, can substantially enhance field emission at nanoscale \cite{PRL1992singleatom}.
It can also be engineered via emitter nanostructuring and has been demonstrated using thin Si/SiO$_2$ \cite{litovchenko1999observation} and carbon \cite{litovchenko2004quantum} layers, as well as silicon nanostructures \cite{johnson2007universal,semenenko2020resonant}. 
Resonant FN tunneling can also be induced by atomically thin dielectrics on the emitter surface \cite{zhou2020theory,henkel2019resonant},
and has been analyzed by solving the one-dimensional Schr\"odinger equation numerically \cite{tan2016theoretical} or analytically \cite{zhou2020theory,zhou2023theoretical}.
More recently, transfer-matrix methods have been used \cite{davidovich2023field}.

Field-emission devices have experienced a revival with the advent of air-channel architectures that conduct current across $\sim$10 nm gaps where electron-molecule collisions are negligible \cite{nirantar2018metal}. Compared with traditional vacuum electronics, such extreme downscaling reduces operating voltages to CMOS-compatible levels \cite{fan2021sub,chang2020field} and enables integration with semiconductor platforms \cite{tang2024beta,sapkota2021ultralow,chang2019vertical,liu2017sic}, opening opportunities for high-speed, robust operation \cite{cao2023future,pradhan2024materials}. Nanoscale electrode geometries further concentrate local fields, yielding large enhancement factors \cite{wang2024enhancing,yamaguchi2011field}. Within this context, resonant field emission offers a pathway to engineered, non-monotonic $I-V$ characteristics for air-channel nanoelectronics.
 
In this Letter, we propose using two-dimensional (2D) conductors to harness and tune resonant tunneling for field emission. 
In materials such as graphene \cite{novoselov2004electric} and emerging 2D metals \cite{zhao2025realization}, out-of-plane electron motion is in the ultra-quantum limit and effectively confined to a single 2D subband. When such a conductor is physisorbed on a bulk metal, the angstrom-scale interfacial gap can support a clean, single resonance in the tunneling probability, yielding pronounced, tunable enhancements in emission near a characteristic field. We develop this concept using noble-metal/graphene heterostructures, which form well-defined angstrom-scale interfaces \cite{pletikosic2009dirac,varykhalov2010effect,wofford2012extraordinary}, and we ground the analysis in ab initio interface parameters and exact solutions of the Schr\"odinger equation for transmission across the junction.

\begin{figure*}
 \includegraphics[width=0.9\textwidth]{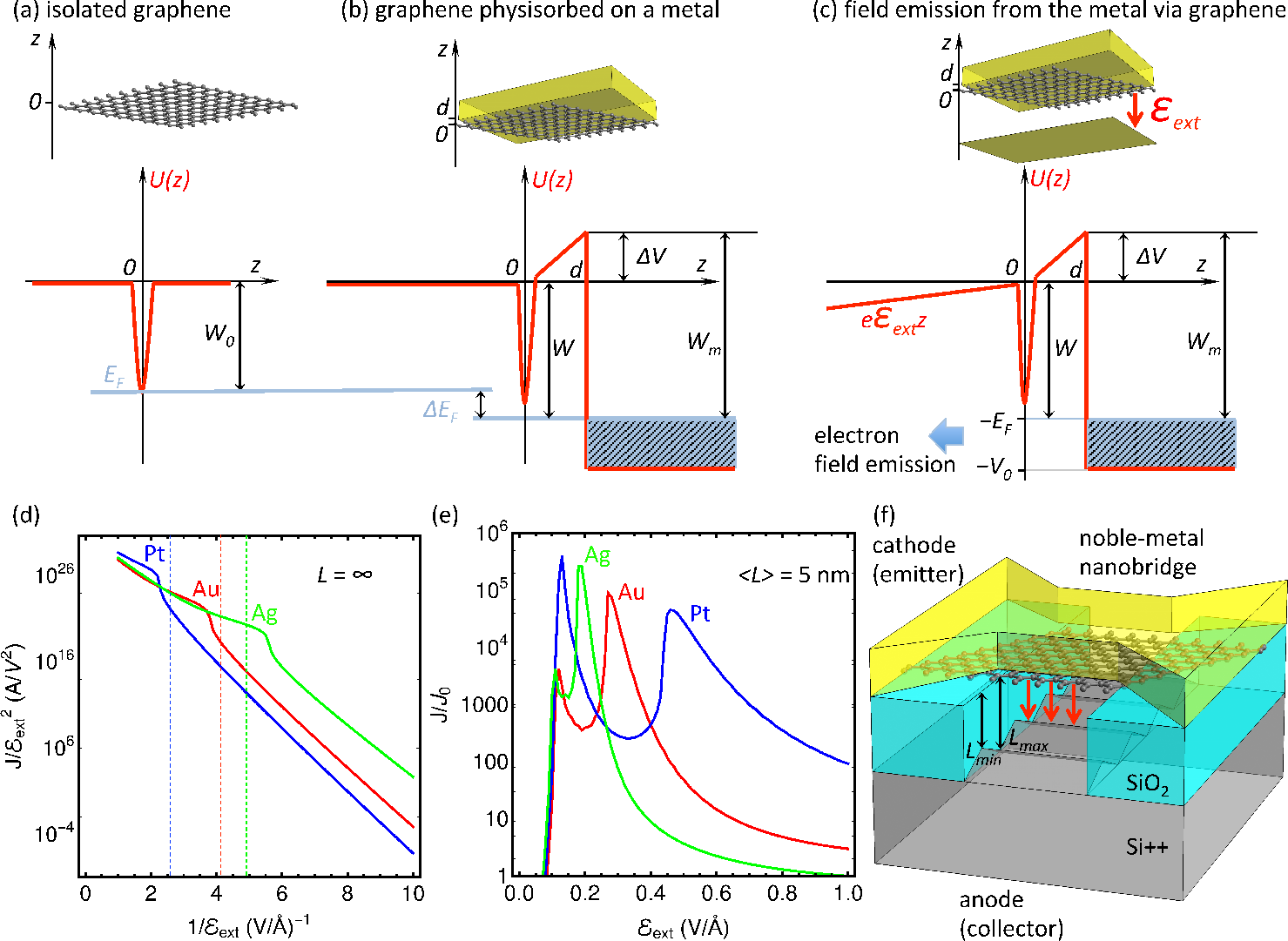}
 \caption{Resonant field-emission by metal-graphene interface.
 (a) Out-of-plane potential profile of isolated graphene exhibits extremely narrow electron confinement, with the work function $W_0=4.55$ eV.
 (b) Upon physisorption on a metal, the work function difference leads to a redistribution of surface charge, 
 shifting the Fermi level by $\Delta E_F$ in graphene and creating an interfacial barrier of height $\Delta V$ and thickness $d$. 
 (c) Field electron emission from a metal covered by graphene is governed by tunneling through a composite potential barrier consisting of
 triangular and trapezoid sections, which supports a resonant state. 
 (d) Fowler-Nordheim representation of the field-emission current density $J$ clearly demonstrates deviation from the conventional 
  linear relation between $\ln(J/{\cal E}_\mathrm{ext}^2)$ and $1/{\cal E}_\mathrm{ext}$
  when $e{\cal E}_\mathrm{ext}\sim \Delta V/d$ (shown by dashed lines) and the resonant level enters the energy window  between $-E_F$ and $-V_0$, see Fig. \ref{fig2}.
  The exact position of the emission maximum also depends on charge doping, which  can be either $n-$ or $p-$type (negative or positive $\Delta E_F$; see Table \ref{tab1}),
  depending on a metal employed (Ag, Au, and Pt). The finite-size channel effects are neglected here (no direct tunneling, see Fig. \ref{fig3}).
 (e) Field emission current density computed for the device shown in (f) and normalized to $J_0$ computed for the same geometry  without graphene.
  The curves are averaged over 100 channels of randomized lengths between $L_\mathrm{min}=4.5$ nm and $L_\mathrm{max}=5.5$ nm to take into account the collector surface roughness. The model neglects the field enhancement factors (e.g. the image-charge effects), which are addressed in Fig. \ref{fig4}.
 (f) Possible realization of the resonance field-emission on a heavily doped Si/SiO$_2$ wafer with a bow-tie gold bridge and SiO$_2$ etched away underneath, see Ref. \cite{tayari2015tailoring}. 
 Doped Si is usually used as a gate contact but if SiO$_2$ is removed completely under the nanobridge, it plays an electron collector role.
 \label{fig1}
 }
\end{figure*}

Figure \ref{fig1} summarizes the central result by contrasting resonant emission with classical FN tunneling and illustrating a vertical noble-metal/graphene nanobridge emitter above a doped-Si collector. We examine Au, Pt, and Ag contacts, focusing on gold because ARPES measurements for graphene on Au \cite{varykhalov2010effect} align closely with the ab initio interface parameters used in our simulations.
We then analyze finite-size effects, including the transition to direct tunneling, surface roughness, and electrode curvature, and finally demonstrate resonant field emission in a coplanar geometry with sharp electrodes that enable strong field enhancement.

Strong out-of-plane confinement in a 2D conductor quenches the electron kinetic energy along the out-of-plane direction, creating a theoretical challenge for describing out-of-plane emission \cite{qin2011analytical,ang2018universal,trushin2018theory,lepetit2021quantum}; see also recent reviews \cite{chan2022field,ang2021physics}. Consequently, the central ingredient in modeling electron emission through a 2D material is the choice of the out-of-plane confining potential. 
A minimal and effective way to model this confinement is a delta-function potential, $-u_0\delta(z)$, which captures the ultra-quantum limit with a single out-of-plane subband, see Fig. \ref{fig1}(a). In this framework, the potential strength $u_0>0$ sets both the work function and the decay length:
$W_0=mu_0^2/(2\hbar^2)$ and $l_0=\hbar^2/(mu_0)$, where $\hbar$ is the reduced Planck constant, $m$ is the free electron mass governing out-of-plane motion.
A full derivation of graphene's electronic states under delta-function confinement is provided in Supporting Information S1.

The functionality of the proposed device relies on graphene physisorbing, rather than chemisorbing, onto a noble-metal surface.
Chemisorption alters graphene's band structure and reduces the graphene-metal separation $d$ to 2--2.5 \AA, whereas physisorption preserves graphene’s electronic structure at larger separations ($d>3$ \AA) \cite{tao2018modeling}. Note that $d$ differs from the AFM-measured graphene-substrate offset,
which includes the full graphene thickness and interfacial contaminants \cite{liu2014giant}. 
High-resolution ARPES confirms physisorption on iridium \cite{pletikosic2009dirac} and gold \cite{varykhalov2010effect}, where graphene retains its gapless Dirac dispersion,
in contrast to nickel \cite{graphene-nickel} and copper \cite{varykhalov2010effect}, which induce stronger hybridization.
Even in the physisorbed regime, however, proximity-induced charge transfer can dope graphene and shift its Fermi level by $\Delta E_F$,
changing the work function to $W=W_0+\Delta E_F$ \cite{dopingPRL2008}.
Crucially, electron transfer does not fully equilibrate the work functions in graphene ($W$) and metal ($W_m$):
an interface dipole layer forms and maintains a finite barrier $\Delta V= W_m-W$ \cite{dopingPRB2009}, 
which separates graphene's 2D states from the bulk metal states.
We model this interfacial barrier as trapezoidal with a constant potential gradient $\Delta V/d$, see Fig. \ref{fig1}(b),
and its parameters are obtained by fitting density-functional theory (DFT) data \cite{dopingPRL2008}
to a phenomenological model \cite{dopingPRB2009}. 
A summary of $W_m$, $W$, $\Delta E_F$, and $\Delta V$ is provided in Table \ref{tab1}.

\begin{table}
\begin{center}
\begin{tabular}{c|c|c|c|c}
Metal & $W_m$ (eV) & $W$ (eV) & $\Delta E_F$ (eV) & $\Delta V/d$ (eV/\AA)   \\
\hline
Ag & 4.92          & 4.24     &   $-0.32$         & 0.20                   \\
Au & 5.54          & 4.74     &   $+0.19$         & 0.24                   \\
Pt & 6.13          & 4.87     &   $+0.33$         & 0.38                    
\end{tabular}
\end{center}
\caption{Work functions of bulk metals ($W_m$) and graphene ($W$) adsorbed on their surfaces, computed via DFT \cite{dopingPRB2009}, with electron transfer captured by
the Fermi energy shift in graphene ($\Delta E_F$). The interfacial barrier is characterized by $\Delta V = W_m - W$ and
an interlayer spacing $d$ of about 3.3\AA, computed within the same ab-initio approach \cite{dopingPRB2009,dopingPRL2008} used to match the band parameters
measured by ARPES \cite{varykhalov2010effect}.
\label{tab1}}
\end{table}

To describe field emission, we introduce the conduction-band depth in the bulk, $V_0-E_F$,
typically about 5 eV in noble metals, and the external field ${\cal E}_\mathrm{ext}$, see Fig. \ref{fig1}(c).
Given the high electron concentration in noble metals, we assume $E_F$ and $\Delta V$ are unaffected by emission.
The emission current is computed from the exact solution of the Schr\"odinger equation for out-of-plane electron motion
in the potential profile shown in Fig.\ref{fig1}(c), with the Hamiltonian $\hat H=  \hbar^2\hat k_z^2/(2m)+U(z)$, where $k_z=-i\partial/\partial z$, and $U(z)$ reads
\begin{eqnarray}
 U(z)= -u_0\delta(z)+
 \left\{\begin{array}{ll}
    e{\cal E}_\mathrm{ext}z, & z<0,\\
   \Delta V z/d,& 0<z<d,\\
   -V_0, & z\geq d;
   \end{array}
   \right.
\end{eqnarray}
where $e$ is the elementary charge.
The solution to the  Schr\"odinger equation can be written as
\begin{eqnarray}
\label{wf}
 \psi(z)=
 \left\{\begin{array}{ll}
    A_1\left[\mathrm{Ai}(\zeta_1)+i\mathrm{Bi}(\zeta_1)\right], & z<0,\\
    A_2 \mathrm{Ai}(\zeta_2)+B_2 \mathrm{Bi}(\zeta_2), & 0<z<d,\\
    A_3 e^{-ikz} +B_3 e^{ikz},& z>d;
   \end{array}
   \right.
\end{eqnarray}
\begin{eqnarray}
&& \zeta_1 =  -\frac{E_k}{e{\cal E}_\mathrm{ext}z_1}+\frac{z}{z_1},\quad z_1=\left(\frac{\hbar^2}{2me{\cal E}_\mathrm{ext}}\right)^{\frac{1}{3}},\\
&& \zeta_2 =  -\frac{E_k d}{\Delta V z_2}+\frac{z}{z_2},\quad z_2=\left(\frac{\hbar^2 d}{2m\Delta V}\right)^{\frac{1}{3}},\\
&& k=\sqrt{\frac{2m}{\hbar^2}\left(V_0 + E_k\right)},
\end{eqnarray}
where $\mathrm{Ai}(\zeta_{1,2})$, $\mathrm{Bi}(\zeta_{1,2})$ are Airy functions. The coefficients are found
by matching the wave function and its derivative at $z=0$ and $z=d$. This results in four equations given by
\begin{eqnarray}
&& \label{e1} A_1\left[\mathrm{Ai}(\zeta_{10})+i\mathrm{Bi}(\zeta_{10})\right] = A_2\mathrm{Ai}(\zeta_{20})+B_2 \mathrm{Bi}(\zeta_{20}),\\
\nonumber &&  \frac{A_1}{z_1}\left[\mathrm{Ai}'(\zeta_{10})+i\mathrm{Bi}'(\zeta_{10})\right]- \frac{1}{z_2}\left[A_2 \mathrm{Ai}'(\zeta_{20})+B_2 \mathrm{Bi}'(\zeta_{20})\right]\\
&&\label{e2} =\frac{2}{l_0}\left[ A_2\mathrm{Ai}(\zeta_{20})+B_2 \mathrm{Bi}(\zeta_{20}) \right],\\
&&\label{e3}  A_2\mathrm{Ai}(\zeta_{2d})+B_2 \mathrm{Bi}(\zeta_{2d})  = A_3 e^{-ikd} +B_3 e^{ikd},\\
&& \nonumber   \frac{1}{z_2}\left[A_2 \mathrm{Ai}'(\zeta_{2d})+B_2 \mathrm{Bi}'(\zeta_{2d})\right]\\
&& \label{e4} =  -ik A_3 e^{-ikd} +ik B_3 e^{ikd},
\end{eqnarray}
where $\mathrm{Ai}'(\zeta_{1,2})$, $\mathrm{Bi}'(\zeta_{1,2})$ are derivatives of the Airy functions with
$\zeta_{10}=\zeta_1(z=0)$, $\zeta_{20}=\zeta_2(z=0)$, $\zeta_{2d}=\zeta_2(z=d)$.
Solving Eqs. (\ref{e1}--\ref{e4}) yields $A_{1,2,3}$ and $B_2$.
The incident, reflected, and transmitted probability current densities are, respectively, given by
\begin{equation}
 j_i=\frac{\hbar k}{m}\lvert B_3 \rvert^2,\quad j_r=\frac{\hbar k}{m}\lvert A_3 \rvert^2, \quad j_t  = \frac{\hbar}{\pi mz_0} \lvert A_1 \rvert^2,
\end{equation}
where we have used the identity
\begin{equation}
 \mathrm{Ai}(\zeta)\mathrm{Bi}'(\zeta)-\mathrm{Ai}'(\zeta)\mathrm{Bi}(\zeta) =1/\pi.
\end{equation}
By convention, we take $k<0$ for the incident current propagating toward $z\to -\infty$, and we set $B_3=1$ for normalization.
The transmission and reflection probabilities then read
\begin{equation}
 T= \frac{j_t}{j_i},\quad R= \frac{j_r}{j_i},
\end{equation}
which satisfy $T+R=1$.

\begin{figure*}
 \includegraphics[width=0.9\textwidth]{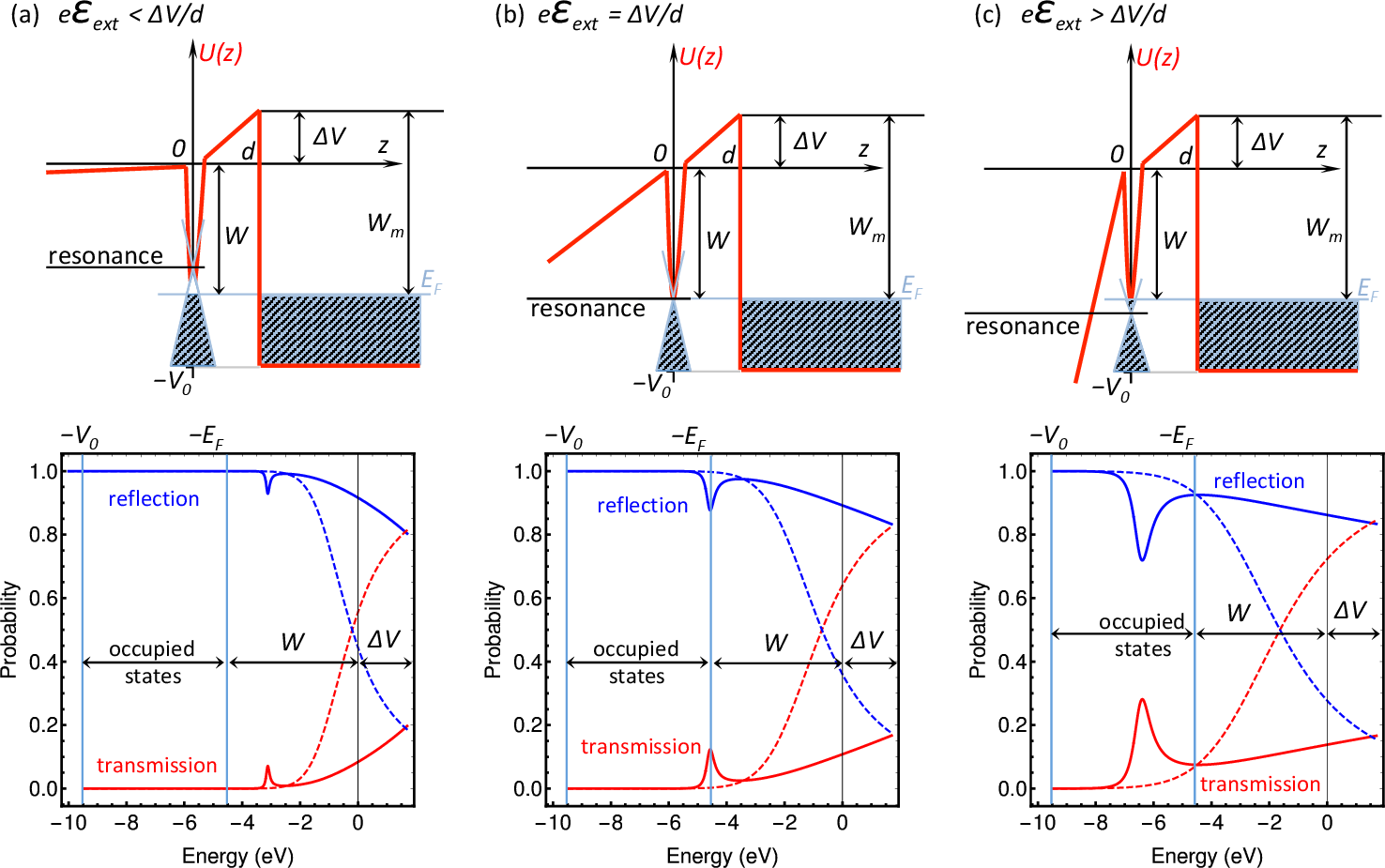}
 \vspace{1cm}
 \caption{Resonant features in transmission (red) and reflection (blue) probabilities with (solid) and without (dashed) graphene.
 Parameters are chosen such to make the resonance clearly visible, $W_m=6.2$ eV, $W=4.5$ eV, $\Delta E_F=0$, $V_0=9.5$ eV,
 $d=0.17$ nm, and they do not correspond to a specific metal.
 (a) For $e{\cal E}_\mathrm{ext} < \Delta V/d$, the resonance lies above the Fermi level; emission from occupied states occurs via resonance broadening.
 (b) For $e{\cal E}_\mathrm{ext} = \Delta V/d$, the resonance matches the Fermi level, producing a dramatic increase in emission.
 (c) For $e{\cal E}_\mathrm{ext} > \Delta V/d$, the resonance falls within the occupied energy window between the metal conduction-band bottom and the Fermi level. At such high fields, however, non-resonant emission without graphene also becomes efficient, reducing $J/J_0$.
 With further increase of ${\cal E}_\mathrm{ext}$ the resonance exit the occupied energy window and the emission current drops to the non-resonant values.
 \label{fig2}
 }
\end{figure*}

\begin{figure*}
 \includegraphics[width=0.9\textwidth]{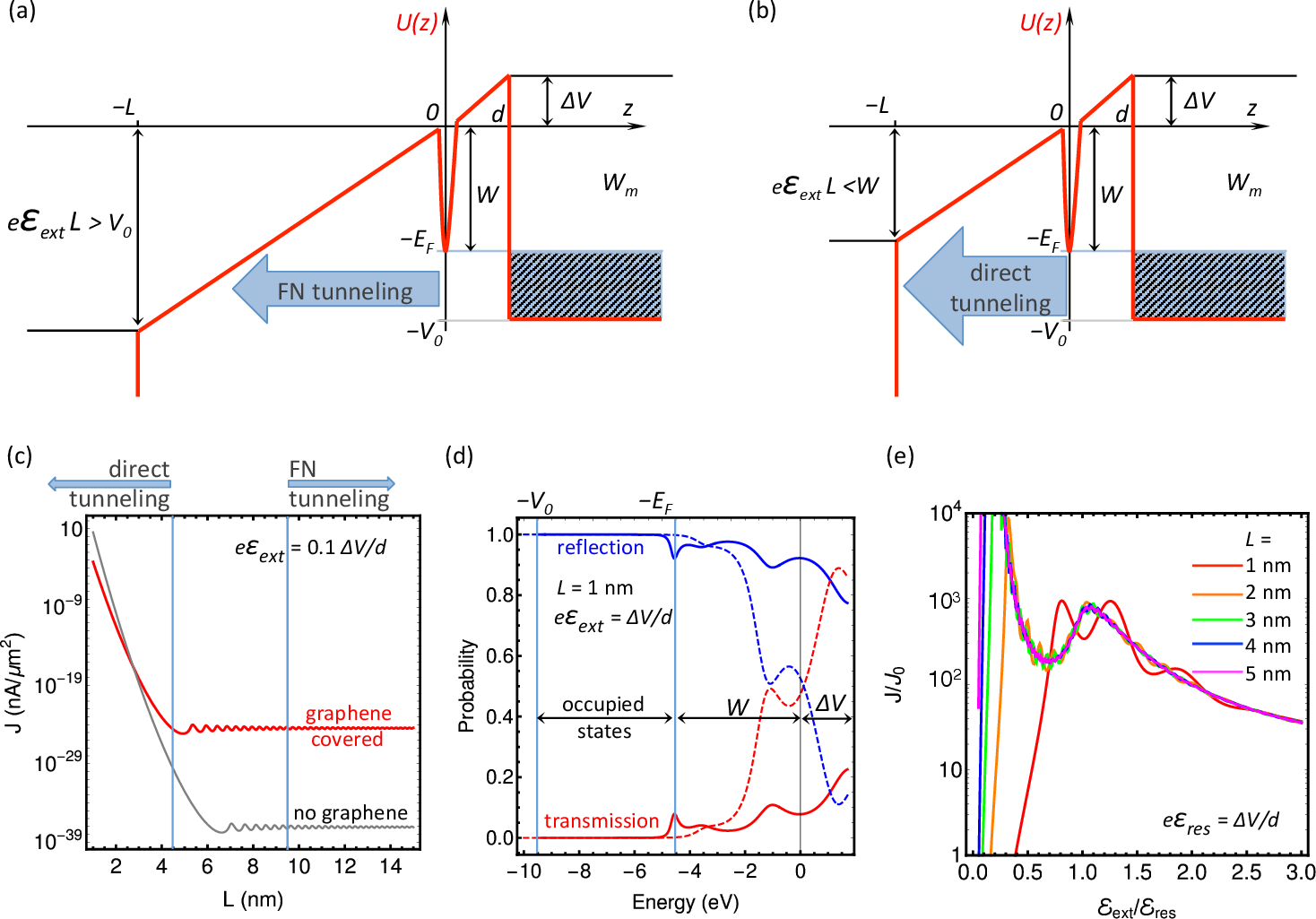}
  \vspace{1cm}
 \caption{Effects of finite channel length $L$. Parameters are the same as in Fig. \ref{fig2} and do not correspond to a specific metal.
 (a,b) Potential profiles for FN  ($e{\cal E}_\mathrm{ext}L > V_0$) and direct tunneling ($e{\cal E}_\mathrm{ext}L<W$) regimes, respectively.
 The intermediate field range $W\leq e{\cal E}_\mathrm{ext}L\leq V_0$ corresponds to a transitional regime in which part of the occupied states tunnel via FN and the remainder via direct tunneling. 
 (c) Distinction between FN and direct tunneling. In direct tunneling, the emission current decreases exponentially
 with increasing $L$,  whereas in the FN regime it is essentially independent of $L$. Moreover, graphene reduces the direct-tunneling current compared with bare metal,
 while enhancing emission in the FN regime. 
 (d) Transmission (red) and reflection (blue) probabilities with (solid) and without (dashed) graphene for a finite channel length of 1 nm, showing additional resonance-like features due to Fabry-Perot interference, cf. Fig. \ref{fig2}(b). 
 (e) Field-emission current ratio $J/J_0$ versus ${\cal E}_\mathrm{ext}/{\cal E}_\mathrm{res}$, computed for different $L$, where
  $e{\cal E}_\mathrm{res}=\Delta V/d$  and $J_0$ is the current for the same channel without graphene.
  In the direct-tunneling regime, the resonance maximum is distorted by backscattering between the electrodes.
 \label{fig3}
 }
\end{figure*}

The field emission current can be written as an integral over $j_t(k)$ and is given by
\begin{equation}
J=2e\int \frac{dk}{2\pi} \int \frac{d^2k_\parallel}{(2\pi)^2} j_t(k),
\end{equation}
where we integrate over in-plane ($k_\parallel$) and out-of-plane ($k>0$) wave vectors. 
The Fermi level sets the integral limits as $k_\parallel \leq  \sqrt{\frac{2m}{\hbar^2}\left(-E_k-E_F\right)}$
and $k \leq  \sqrt{\frac{2m}{\hbar^2}\left(V_0-E_F\right)}$.
Here, $E_k=-V_0+\hbar^2 k^2/(2m)$ and $E_F=W_0+\Delta E_F$, see Fig. \ref{fig1}(a-c) and Table \ref{tab1}.
The electron effective mass ($m$) and conduction band depth $V_0-E_F\sim 5$ eV are taken to be the same for all the metals considered,
as their particular values have negligible effects on the resonant emission. Taking the integral over 
$k_\parallel$ and changing the variable $dk$ to $dE_k$ (see Supporting Information S2) we arrive at the final expression for $J$ given by
\begin{eqnarray}
 \nonumber J &=& \frac{e}{8\pi^2}\left(\frac{2m}{\hbar^2}\right)^{\frac{3}{2}}\int\limits_{-V_0}^{-E_F} dE_k \frac{-E_F-E_k}{\sqrt{V_0+E_k}}\\
 && \times j_t\left(\sqrt{\frac{2m}{\hbar^2}\left(V_0+E_k\right)}\right).
\end{eqnarray}

Figure \ref{fig1}(d) shows the emission current density $J$ for three noble-metal/graphene heterostructures (Ag, Au, and Pt),
plotted on a log-scale as $J/{\cal E}_\mathrm{ext}^2$ versus $1/{\cal E}_\mathrm{ext}$. 
While a linear dependence is expected in the FN regime (see Supporting Information S2), the response here is nonmonotonic, exhibiting a single resonance near
$e{\cal E}_\mathrm{ext}\sim \Delta V/d$. 

Figure \ref{fig1}(e) presents the resonant field emission for the same noble-metal/graphene heterostructures as in Fig. \ref{fig1}(d), 
now evaluated for the specific device shown in Fig. \ref{fig1}(f). 
The current density $J$ is normalized to $J_0$, computed for the corresponding metal under the same conditions but without graphene. 
The simulations incorporate finite-size effects (an average emitter-collector distance separation $\langle L \rangle=5$ nm),
as well as collector surface unevenness inevitable upon etching SiO$_2$ away \cite{tayari2015tailoring} (see Supporting Information S3 for details). 
A clear, orders-of-magnitude enhancement in current is observed near the resonant field $e{\cal E}_\mathrm{ext}\sim \Delta V/d$.

Figure \ref{fig2}(a-c) illustrates resonant field emission through graphene under the simplifying assumption that graphene's work function $W$
does not change upon contact with a metal and remains equal to that of pristine (intrinsic) isolated graphene. 
This is the assumption made in only Fig. \ref{fig2} and Fig. \ref{fig3} to demonstrate the basics of the model.
The interfacial barrier $\Delta V=W_m- W>0$ is thus taken to be constant. The band-crossing (charge-neutrality) point
can be shifted upward by $\Delta V$ or downward by an external field ${\cal E}_\mathrm{ext}$, as modeled in Supporting Information S1.
In the limit  $e{\cal E}_\mathrm{ext}\ll \Delta V/d$, the band-crossing point is pushed to its highest energy set by $\Delta V$ and $d$;
in the opposite limit $e{\cal E}_\mathrm{ext}\gg \Delta V/d$, it is lowered and governed by ${\cal E}_\mathrm{ext}$.
When the external and built-in fields balance ($e{\cal E}_\mathrm{ext}= \Delta V/d$), 
the situation reduces to intrinsic graphene, with the Fermi level intersecting the band-crossing point.

The lower panels of Fig. \ref{fig2}(a-c) show that the resonance peak in the transmission probability can be tuned by ${\cal E}_\mathrm{ext}$
in the same way as the band shifts: the peak moves to lower energy with increasing ${\cal E}_\mathrm{ext}$, 
and the shift becomes noticeable when $e{\cal E}_\mathrm{ext}$ is of the order of $\Delta V/d$.
The parameter values given in the figure caption are chosen to clearly demonstrate the resonance peak; unlike Fig. \ref{fig1}, they do not correspond to a specific metal. In realistic models, $\Delta V$ is typically much smaller and $d$ somewhat larger, which results in a much narrow resonance that is harder to resolve in the plot. Increasing ${\cal E}_\mathrm{ext}$ not only shifts the resonance but also broadens it, because the quasibound state becomes short-lived, leading to smeared resonance features. If graphene is removed, then both transmission and reflection probabilities become featureless, see Supporting Information S2.

The transmission probability itself does not depend on the Fermi level, however, the Fermi level determines whether a given transmission mode contributes to field emission. In the simplified model shown in Fig. \ref{fig2}(a-c), the resonant level aligns with the Fermi level at $e{\cal E}_\mathrm{ext}= \Delta V/d$ and begins to contribute to electron emission strongly that manifests as an exponential burst in the emission current density compared with the case without graphene. 
In real metals, graphene-metal heterostructures are characterized not only by the interfacial barrier height $\Delta V$ but also the Fermi-level shift $\Delta E_F$ (see Table \ref{tab1}). For silver contacts, $\Delta E_F<0$, which lowers graphene's work function. As a consequence, the resonant mode drops below the Fermi level and starts contributing to emission after $1/e{\cal E}_\mathrm{ext}$ reaches $d/\Delta V$, see the green curve in Fig. \ref{fig1}(d). In contrast, gold and platinum induce $p-$type doping in graphene ($\Delta E_F>0$), increasing its work function, so the resonant mode requires $1/e{\cal E}_\mathrm{ext}<d/\Delta V$ to contribute, see the red and blue curves in Fig. \ref{fig1}(d).
Because $\Delta V$ is nearly twice as large for platinum as for gold or silver, the corresponding critical $1/{\cal E}_\mathrm{ext}$ is also about twice as low.



To elucidate the role of channel length, we contrast FN and direct-tunneling regimes, see Fig. \ref{fig3}(a,b).
The solution follows the classical FN procedure, augmented by an additional boundary condition at $z = -L$ (see Supporting Information S3).
In the FN regime, the barrier shape is set by the external field, so electrode geometry has little effect on the tunneling probability. This is a well-known property that makes FN tunneling robust, albeit requiring strong fields. The emission starts transitioning from FN to direct tunneling when the gap length $L$ becomes smaller
than $W/(e{\cal E}_\mathrm{ext})$. The triangular barrier is then truncated to a trapezoidal one, with the truncation set by $L$. 
While the FN current is essentially independent of $L$, in the direct-tunneling regime the current decreases exponentially with increasing $L$, see Fig. \ref{fig3}(c).
In addition, backscattering between the contacts produces Fabry-Perot-like interference in the transmission, see Fig. \ref{fig3}(d), leading to oscillatory features in the emission current that can partially mask the primary resonance at shorter $L$, see Fig. \ref{fig3}(e). 
Finally, the ratio $J/J_0$ drops in the direct-tunneling regime, which dominates at weaker fields.

With the length-dependent model in hand, we can simulate realistic nanometer-scale devices. 
Arbitrary electrode pairs can be treated by partitioning each surface into small planar facets (Supporting Information S3), yielding a set of emission channels, each defined by a local gap length $L_n$ and by the local field component normal to the surface. Surface roughness, such as that indicated for the anode in Fig. \ref{fig1}(e), can be incorporated by randomizing $L_n$ within $[L_\mathrm{min}, L_\mathrm{max}]$. Averaging over on the order of a hundred channels produces a smooth $I-V$ curve that retains the prominent resonance feature.

Within the FN regime, however, it is useful to distinguish smooth from sharp roughness. For smooth roughness, where graphene conforms to the metal and the graphene-metal separation $d$ remains approximately constant, the resonance is preserved. For sharp roughness, graphene may sag locally or partially delaminate, producing spatial variations in $d$. In this case, each channel resonates at a slightly different field, yielding a broadened yet still identifiable resonance peak, provided $d$ fluctuates within 1--2\,\AA, see Supporting Information S3.

\begin{figure}
 \includegraphics[width=0.8\columnwidth]{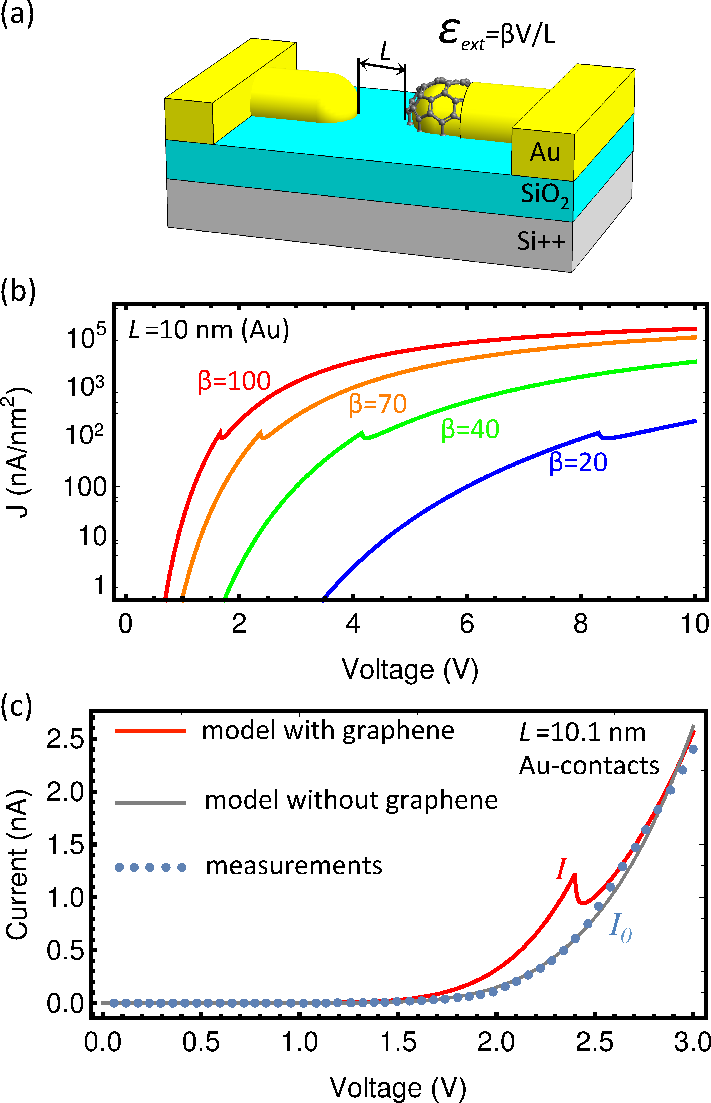}
 \caption{Mapping the model to a coplanar device geometry with pointy contacts.
 (a) Possible schematics of a resonant field-emission device. The graphene layer is not to scale.
 The doped Si/SiO$_2$ substrate can be used for electrostatic gating \cite{nirantar2018metal}. 
 (b) Current density versus applied voltage for a gold/graphene emitter, computed for different $\beta$ values; other parameters are taken from Table \ref{tab1}.
 (c) Current-voltage characteristics for a specific device with $L=10.1$ nm and $\beta=70$.
 Gray curve is obtained by fitting the field-emission model to experimental data (blue dots) for bare gold electrodes \cite{wang2024enhancing}.
 Red curve is computed using the same geometric parameters but with a graphene-coated emitter. The model predicts a resonance near $V \sim 2.3$ V.
 \label{fig4}
 }
\end{figure}

Figure \ref{fig4} examines the field-enhancement factor $\beta$ \cite{leonard2008field,chan2022field},
which captures image-charge and geometric effects, and is specific to each device. 
Because $\beta$ is especially pronounced for pointy contacts \cite{yamaguchi2011field}, we adopt a coplanar geometry with sharp electrodes, see Fig. \ref{fig4}(a),
similar to recent experimental setups \cite{nirantar2018metal,wang2024enhancing}.
The applied voltage $V$ relates to the external field as ${\cal E}_\mathrm{ext}=\beta V/L$, where $L$ is the electrode separation.
Figure \ref{fig4}(b) shows resonant field-emission current density versus $V$ for typical $\beta$ and $L$ values,
demonstrating that the emission can be tuned to resonance at voltages of a few volts.

To address a finite curvature of the emitter's tip, each surface is partitioned into infinitesimal planar facets, as in the roughness modeling of Fig. \ref{fig1}(e,f).
With finite curvature, the local field direction varies across the surface; this is included by projecting ${\cal E}_\mathrm{ext}$ onto each facet's normal.
Using this approach, we simulate electrodes of radius 10 nm for several characteristic separations (Supporting Information S3). 
The results show that emission is dominated by the central channel at the apex, where the local field aligns with the shortest tunneling path,
consistent with established field-emission theory. 

We determine $\beta$ by fitting recent experimental data on field-emission from gold nanorods \cite{wang2024enhancing}, writing $I_0=AJ_0$, where $A$ is an effective emission area adjusted to match the measurements. Keeping the same geometry, we then predict the $I-V$ characteristics for a graphene-coated emitter.
Figure \ref{fig4}(c) shows that $I_0$ reproduces the field-emission data for bare gold nanorodes with $L=10.1$ nm. For this relatively large gap,
low-voltage direct tunneling is much weaker than high-voltage FN emission, so the latter dominates in $I_0(V)$. Using the same $\beta$ and $A$ as for $I_0$, the computed $I=AJ$
(red curve) reveals substantial enhancement, and, critically, a resonant peak near $V\sim 2.3$ V. 
A region of negative differential conductance appears between 2.3 and 2.5 V. At higher voltages, $I_0(V)$ and $I(V)$ converge, as non‑resonant emission becomes comparable to the resonant channel.

To conclude, two device geometries emerge from this work, each with distinct trade-offs. The suspended nanobridge leverages established fabrication \cite{tayari2015tailoring} and fully exploits the spatially uniform resonant emission provided by a flat emitter. 
Its main limitation is electrostatic control: with the doped-Si collector serving as the anode, conventional back-gating is not available. 
This could be mitigated by side-gate electrodes, or a split collector \cite{youh2015flat,ji2020study}.
In contrast, the coplanar pointed electrodes readily support gating, as demonstrated experimentally \cite{nirantar2018metal},
and benefit from large field-enhancement factors. However, 
conformal graphene coating is more challenging, though feasible for related architectures (e.g., nanoparticle arrays \cite{osvath2015structure}).

Beyond monolayer graphene, the model naturally generalizes to multilayers by replacing a single $\delta$-function with a stack of $\delta$-potentials.
A double-layer example (Supporting Information S1) reproduces bilayer-graphene band splitting \cite{mccann2007low}, as well as field- or proximity-induced symmetry breaking.
Qualitatively, each additional layer introduces an extra resonance and inter-resonance interference akin to Fig. \ref{fig3}(d).
This provides added tunability but increases the risk of spectral overlap and inhomogeneous broadening. Practically, we expect clean, resolvable features for one to two layers, provided $d$ fluctuations remain within 1--2\,\AA.

Alternative 2D coatings merit consideration. In contrast to graphene, semiconducting 2D transition metal dichalcogenides (2D TMDC), such as MoS$_2$ or WSe$_2$,
exhibit substantial hybridization on noble metals \cite{nagireddy2025electronic}, which invalidates a simple trapezoidal interfacial barrier and complicates out-of-plane confinement. That said, hybridization is interface-dependent: passivation layers (e.g., by an hBN spacer) or specific metals can restore weak coupling and a clean barrier. 
However, 2D TMDC thickness (up to three atomic layers) also challenge the single-$\delta$ potential approximation.
More complex stacks, such as 2D metals encapsulated by pair of 2D TMDCs layers \cite{zhao2025realization,wang2025pressure,ren2025metallene},
or Janus 2D TMDCs \cite{picker2025atomic,xu2025interface,pan2025layer}
offer additional tuning knobs (built-in dipoles, internal polarization) but introduce intricate interfacial electronic structure
that may obscure a single sharp resonance without careful engineering.

In summary, noble-metal/graphene heterostructures offer a practical route to resonant field emission with angstrom-scale interfaces, minimal hybridization, and a single out-of-plane subband --- all conducive to a strong, tunable resonance at technologically reasonable voltages. 
The resulting non-monotonic $I-V$  characteristics with a pronounced negative differential conductance create opportunities for compact oscillators and other 
electronic components that remain robust at elevated temperatures. Near-term priorities include: (i) scalable graphene coating of pointed electrodes, 
(ii) precise control and metrology of $d$, and (iii) validation of resonance position and linewidth versus geometry and environment.
These steps will clarify the performance envelope and integration pathways alongside air-channel nanoelectronics.

I thank Davit Ghazaryan for discussions.
This research is supported by the Singapore Ministry of Education Research Centre of Excellence award to the Institute for Functional Intelligent Materials
(I-FIM, Project No. EDUNC-33-18-279-V12).

\bibliography{resonantFE.bib}

\appendix

\setcounter{equation}{0}
\renewcommand{\theequation}{S\arabic{equation}}
\setcounter{figure}{0}
\renewcommand{\thefigure}{S\arabic{figure}}

\begin{center}
SUPPORTING INFORMATION
\end{center}

\section{S1. Out-of-plane confinement for electrons in graphene}


In general, electrons in a 2D conductor can be described by the Hamiltonian 
\begin{equation}
 \hat H=\hat H_\parallel + \hat H_\perp,
\end{equation}
where $\hat H_\parallel$ and $\hat H_\perp$ describe in-plane and out-of-plane motion, respectively.

The in-plane Hamiltonian is typically derived from a tight-binding model that exploits lattice periodicity. 
For graphene, the low-energy tight-binding model on a honeycomb lattice with two interpenetrating sublattices 
yields the effective 2$\times$2 Hamiltonian given by
\begin{equation}
 \hat H_\parallel =  \hbar v \left( 
\begin{array}{cc}
0 & \hat k_x - i \hat k_y\\
\hat k_x + i \hat k_y & 0
 \end{array}
 \right),
\end{equation}
where $\hat k_{x,y}$ are the in-plane components of the electron wavevector, $v$ is the tight-binding parameter (the effective Dirac velocity), and $\hbar$ is the Planck constant. The spectrum of $\hat H_\parallel$ is linear for both conduction and valence bands.

\begin{figure*}
 \includegraphics[width=0.6\textwidth]{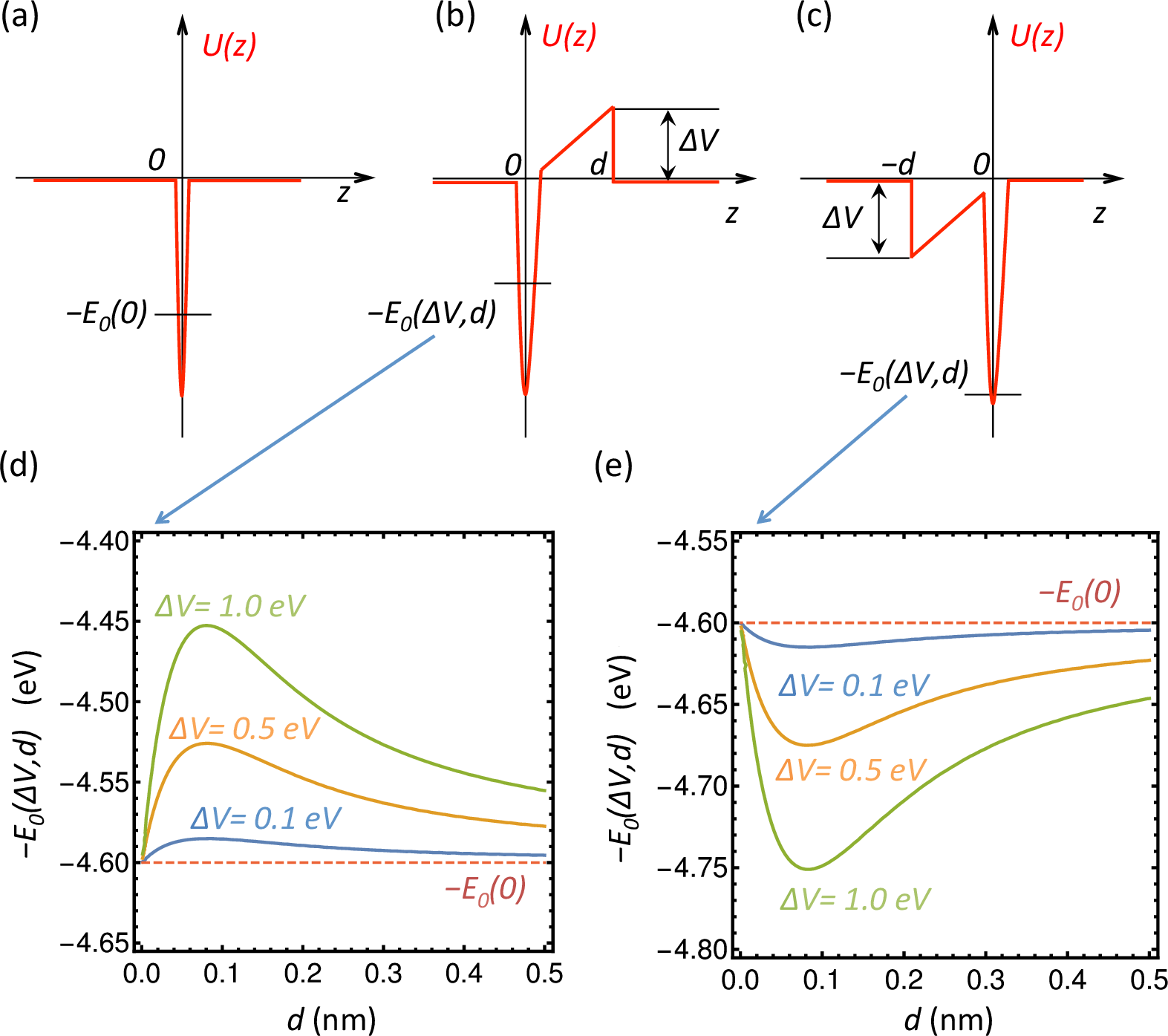}
 \vspace{1.5cm}
 \caption{Out-of-plane confinement of electrons in graphene modeled by a delta-function potential. 
 (a) Electrons in an isolated delta-function potential have a single bound state at energy $-E_0$ corresponding to the difference between 
 the vacuum level and the charge-neutrality (band-crossing) point in intrinsic (undoped) graphene. 
 (b,c) An applied electrostatic potential shifts $-E_0$ up or down depending on its sign.
 The potential gradient $\Delta V/d$ may arise from a built-in field at the graphene-metal interface or from an external field.
 (d,e) Solutions of Eqs. (\ref{DV1}) and (\ref{DV2}) showing the shift of $-E_0$ for different $\Delta V$ and $d$. 
 The shift is maximized at $d\sim l_0$ of about an angstrom, reaching a few tens of meV for a potential drop on the order of $0.1$ eV. 
 \label{figS1}
}
\end{figure*}

The out-of-plane term is less straightforward because there is no lattice periodicity along $z$.
Instead, electrons experience very strong confinement leaving only the lowest quantized subband in a quantum well.
A natural model for this situation is a delta-function confinement that supports a single bound state along $z$, see Fig. \ref{figS1}(a).
The out-of-plane Hamiltonian then reads
\begin{equation}
 \hat H_\perp = \frac{\hbar^2\hat k_z^2}{2m_0}- u_0 \delta(z),
\end{equation}
where $u_0$ is the confinement parameter (determined below), and $m_0$ is the free electron mass.

The eigen-states of $\hat H$ can then be written as
\begin{equation}
 \psi_0(x,y,z)=\frac{1}{L\sqrt{l_0}} e^{ik_x x +ik_y y-|z|/l_0}\left(
\begin{array}{c}
1\\
\pm e^{i\theta}
 \end{array}
 \right),
\end{equation}
where $\tan\theta=k_y/k_x$,
and the eigen-values are given by
\begin{equation}
\label{Ek}
E_\pm= \pm \hbar v k_\parallel - E_0,
\end{equation}
where $k_\parallel=\sqrt{k_x^2+k_y^2}$, and $E_0=m_0u_0^2/(2\hbar^2)$ is the energy difference between the vacuum level and the charge neutrality point.
The typical value of $E_0$ is about $4.6$ eV, which corresponds to the work function of pristine (intrinsic) graphene.
Equivalently, $E_0=\hbar^2/(2m_0l_0^2)$, with a characteristic confinement length $l_0=\hbar^2/(m_0 u_0)$ of about an angstrom.
If graphene is doped, the work function becomes $W=E_0+\Delta E_F$, where $\Delta E_F$ is the Fermi level shift
($\Delta E_F<0$ for $n-$type and $\Delta E_F>0$ for $p-$type doping).

\subsection{Delta-shaped confinement in an external field}

Applying a uniform out-of-plane field ${\cal E}_\mathrm{ext}$ from $z=-\infty$ to $z=+\infty$ does not make much
sense, as it renders the electron motion unbounded and eliminates bound states. 
Instead we introduce a potential drop $\Delta V$ at a distance $z=d$ from the origin point, see Fig. \ref{figS1}(b).
The electric field is then taken to be constant, $\Delta V/d$, for $0<z<d$, and to vanish outside this region.
The out-of-plane wave function can be written as
\begin{eqnarray}
 \psi(z)=
 \left\{\begin{array}{ll}
    A_1 e^{\kappa z}, & z<0,\\
    A_2 \mathrm{Ai}(\zeta)+B_2 \mathrm{Bi}(\zeta), & 0<z<d,\\
    A_3 e^{-\kappa z},& z>d;
   \end{array}
   \right.
\end{eqnarray}
where $\kappa=\sqrt{-2m_0 E_\kappa/\hbar^2}$, $\mathrm{Ai}(\zeta)$, $\mathrm{Bi}(\zeta)$ are the Airy functions of $\zeta$ given by
\begin{equation}
\zeta=-\frac{E_\kappa d}{\Delta V z_0}+\frac{z}{z_0}, 
\label{zeta}
\end{equation}
\begin{equation}
z_0=\left(\frac{\hbar^2 d}{2m_0 \Delta V}\right)^{\frac{1}{3}},
\label{zeta0}
\end{equation}
and $E_\kappa=-E_0$ is the new bound-state energy to be found.
The continuity conditions for $\psi(z)$ and $\psi'_z(z)$ result in
the following set of four equations
\begin{equation}
 \left\{\begin{array}{ll}
    A_1 = A_2\mathrm{Ai}(\zeta_0)+B_2 \mathrm{Bi}(\zeta_0),\\
     A_1 \kappa - \frac{1}{z_0}\left[A_2 \mathrm{Ai}'(\zeta_0)+B_2 \mathrm{Bi}'(\zeta_0)\right]=\frac{2}{l_0} A_1,\\
    A_2\mathrm{Ai}(\zeta_d)+B_2 \mathrm{Bi}(\zeta_d)  = A_3 e^{-\kappa d},\\
    \frac{1}{z_0}\left[A_2 \mathrm{Ai}'(\zeta_d)+B_2 \mathrm{Bi}'(\zeta_d)\right]= -A_3\kappa e^{-\kappa d},
   \end{array}
   \right.
   \label{set4}
\end{equation}
where $\zeta_{0,d}$ are given by $\zeta$ taken at $z=0,d$.
The set of equations determines the coefficients $A_{1,2,3}$ and $B_2$, as well as the binding energy $E_0$.
Excluding the coefficients $A_1$ and $A_3$ we obtain a set of two equations given by
\begin{eqnarray}
 \nonumber &&  A_2\left[\kappa \mathrm{Ai}(\zeta_0) - \frac{1}{z_0}\mathrm{Ai}'(\zeta_0)-\frac{2}{l_0} \mathrm{Ai}(\zeta_0)\right]\\
 && +B_2 \left[\kappa \mathrm{Bi}(\zeta_0) - \frac{1}{z_0}\mathrm{Bi}'(\zeta_0)-\frac{2}{l_0} \mathrm{Bi}(\zeta_0)\right]= 0,\\
 \nonumber &&   A_2\left[\kappa \mathrm{Ai}(\zeta_d)+ \frac{1}{z_0}\mathrm{Ai}'(\zeta_d)\right] \\
 && +B_2 \left[\kappa \mathrm{Bi}(\zeta_d)+ \frac{1}{z_0}\mathrm{Bi}'(\zeta_d)\right] = 0.
   \label{set2}
\end{eqnarray}
The set of equations has a non-trivial solution when the determinant vanishes, hence, the eigen-value equation reads
\begin{eqnarray}
\nonumber && \left[\kappa \mathrm{Ai}(\zeta_0) - \frac{1}{z_0}\mathrm{Ai}'(\zeta_0)-\frac{2}{l_0} \mathrm{Ai}(\zeta_0)\right]\left[\kappa \mathrm{Bi}(\zeta_d)+ \frac{1}{z_0}\mathrm{Bi}'(\zeta_d)\right] \\
\nonumber && - \left[\kappa \mathrm{Bi}(\zeta_0) - \frac{1}{z_0}\mathrm{Bi}'(\zeta_0)-\frac{2}{l_0} \mathrm{Bi}(\zeta_0)\right]\\
 && \times \left[\kappa \mathrm{Ai}(\zeta_d)+ \frac{1}{z_0}\mathrm{Ai}'(\zeta_d)\right] = 0.
 \label{DV1}
\end{eqnarray}

The case of a negative potential drop at $z=-d$ shown in Fig. \ref{figS1}(c) can be considered in a similar way.
Note that the potential gradient is still the same as in Fig. \ref{figS1}(b) and given by $\Delta V/d$. The wave function then reads
\begin{eqnarray}
 \psi(z)=
 \left\{\begin{array}{ll}
    A_1 e^{\kappa z}, & z<-d,\\
    A_2 \mathrm{Ai}(\zeta)+B_2 \mathrm{Bi}(\zeta), & -d<z<0,\\
    A_3 e^{-\kappa z},& z>0;
   \end{array}
   \right.
\end{eqnarray}
and the boundary conditions lead to the following set of equations
\begin{equation}
 \left\{\begin{array}{ll}
    A_1 e^{-\kappa d} = A_2\mathrm{Ai}(\zeta_{\bar d})+B_2 \mathrm{Bi}(\zeta_{\bar d}),\\
     A_1 \kappa e^{-\kappa d} = \frac{1}{z_0}\left[A_2 \mathrm{Ai}'(\zeta_{\bar d})+B_2 \mathrm{Bi}'(\zeta_{\bar d})\right],\\
    A_2\mathrm{Ai}(\zeta_0)+B_2 \mathrm{Bi}(\zeta_0)  = A_3,\\
    \frac{1}{z_0}\left[A_2 \mathrm{Ai}'(\zeta_0)+B_2 \mathrm{Bi}'(\zeta_0)\right]+A_3\kappa =\frac{2}{l_0} A_3,
   \end{array}
   \right.
\end{equation}
where $\zeta_{\bar d}$ is given by $\zeta$ taken at $z=-d$.
The set can be simplified as follows
\begin{eqnarray}
 \nonumber &&  A_2\left[\kappa \mathrm{Ai}(\zeta_{\bar d}) - \frac{1}{z_0}\mathrm{Ai}'(\zeta_{\bar d})\right]\\
 && +B_2 \left[\kappa \mathrm{Bi}(\zeta_{\bar d}) - \frac{1}{z_0}\mathrm{Bi}'(\zeta_{\bar d})\right]= 0,\\
 \nonumber &&   A_2\left[\frac{2}{l_0} \mathrm{Ai}(\zeta_0) - \frac{1}{z_0}\mathrm{Ai}'(\zeta_0)- \kappa \mathrm{Ai}(\zeta_0)\right] \\
 && +B_2 \left[\frac{2}{l_0} \mathrm{Bi}(\zeta_0)- \frac{1}{z_0}\mathrm{Bi}'(\zeta_0)-\kappa \mathrm{Bi}(\zeta_0)\right] = 0.
\end{eqnarray}
The eigen-value equation reads
\begin{eqnarray}
\nonumber && \left[\kappa \mathrm{Ai}(\zeta_{\bar d}) - \frac{1}{z_0}\mathrm{Ai}'(\zeta_{\bar d})\right]\left[\frac{2}{l_0} \mathrm{Bi}(\zeta_0)- \frac{1}{z_0}\mathrm{Bi}'(\zeta_0)-\kappa \mathrm{Bi}(\zeta_0)\right] \\
\nonumber && - \left[\kappa \mathrm{Bi}(\zeta_{\bar d}) - \frac{1}{z_0}\mathrm{Bi}'(\zeta_{\bar d})\right]\\
 && \times \left[\frac{2}{l_0} \mathrm{Ai}(\zeta_0) - \frac{1}{z_0}\mathrm{Ai}'(\zeta_0)- \kappa \mathrm{Ai}(\zeta_0)\right] = 0.
 \label{DV2}
\end{eqnarray}

The solutions of Eqs. (\ref{DV1},\ref{DV2}) are shown in Fig. \ref{figS1}(d,e), respectively.
The potential drop $\Delta V$ in Fig. \ref{figS1}(d) can be interpreted as a built-in potential at a graphene-metal interface.
For a graphene-metal spacing of a few angstroms, a built-in potential of 1 eV can raise the band-crossing point toward the vacuum level by about 100 meV.
If the Fermi level is fixed, then graphene becomes $p$-doped. Figure \ref{figS1}(e) illustrates the effect of an external electric field, ${\cal E}_\mathrm{ext}=\Delta V/d$,
on the band-crossing point shift. Here, the field shifts the the band-crossing point downward, and graphene becomes $n$-doped.
In this case, additional bound states may appear in the small triangular quantum well adjacent to the main delta-function confinement, but they occur at a different energy scale and are not shown. In both cases, the energy shift vanishes at $d\gg l_0$.

\begin{figure*}
 \includegraphics[width=0.8\textwidth]{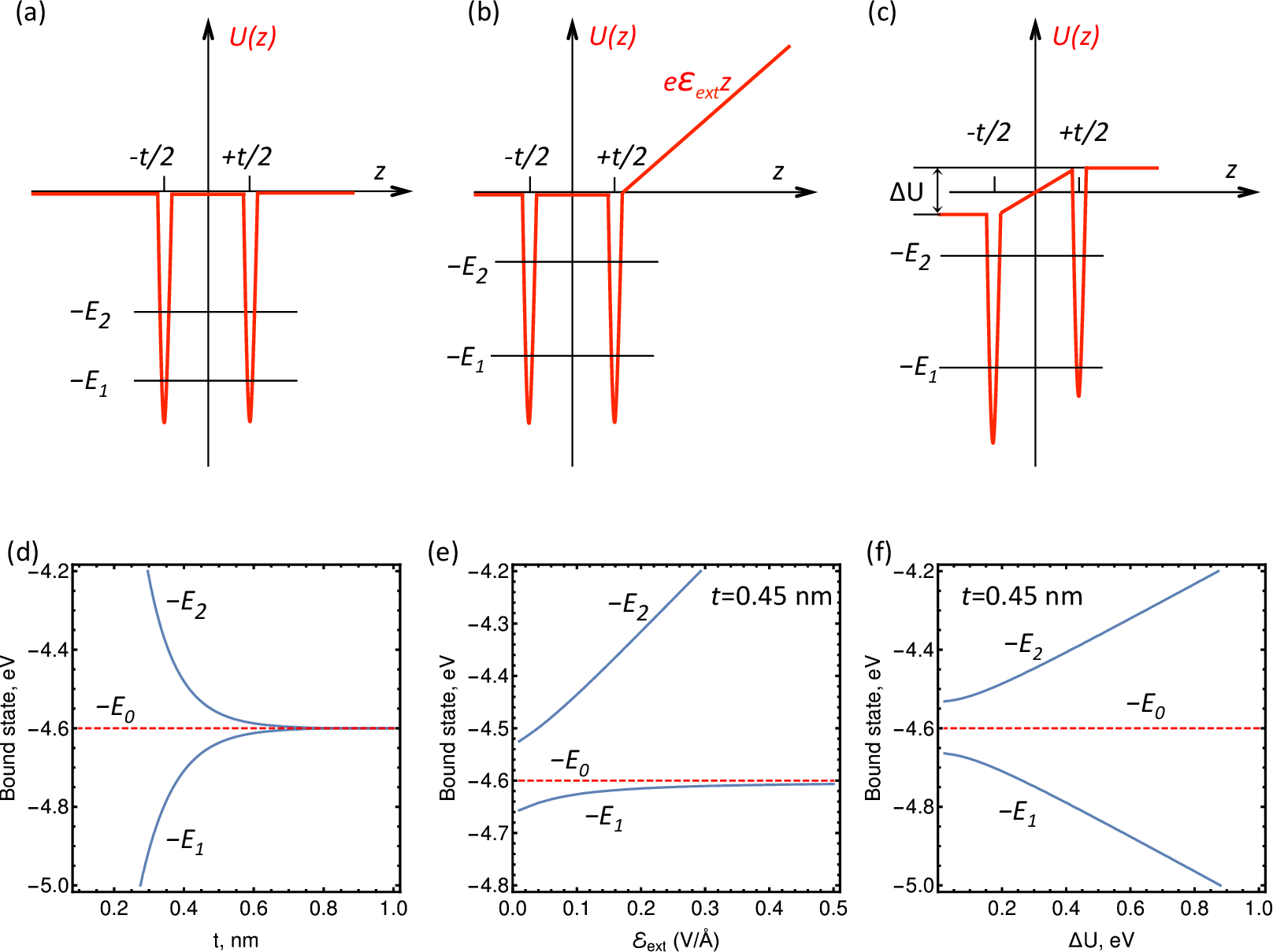}
 \vspace{4cm}
 \caption{Out-of-plane confinement of electrons in a double layer modeled by a pair of delta-function potentials.
 (a) An isolated double-delta potential supports two bound states at energies $-E_1$ and $-E_2$ corresponding 
 to the even and odd parity, respectively.
 (b) An external field applied to one side of the double layer models electrostatic gating or the proximity effect of a metal.
 (c) Internal polarization of the double layer is included via a potential drop $\Delta U$.
 (d) As the interlayer separation $t$ decreases, the bound-state energy splitting $E_1-E_2$ increases
 reaching a few hundred meV at $t$ of a few angstroms, comparable to band splittings in bilayer graphene within a four-band model. 
 Solution is given by Eq. (\ref{eigen2dd}).
 (e) An applied field ${\cal E}_\mathrm{ext}$ drives the lower bound state $E_1$ toward the bound state of an isolated single-delta potential, while
  $E_2$  moves toward the continuum. Solution is given by Eq. (\ref{eigen2ddbias}).
 (f) Interlayer polarization enhances the splitting $E_1-E_2$.  Solution is given by Eq. (\ref{eigenpol}).
 \label{figS2}
}
\end{figure*}

\subsection{Double-layer confinement in an external field}

Let us consider the double-layer confinement, when the out-of-plane electron motion
is limited by a pair of delta-functions, Fig. \ref{figS2}(a). The wave function can be written as
\begin{eqnarray}
 \psi(z)=
 \left\{\begin{array}{ll}
    A_1 e^{\kappa z}, & z<-\frac{t}{2},\\
    A_2 e^{\kappa z}+B_2 e^{-\kappa z}, & -\frac{t}{2}<z<\frac{t}{2},\\
    A_3 e^{-\kappa z},& z>\frac{t}{2};
   \end{array}
   \right.
\end{eqnarray}
where $t$ is the interlayer separation.
Imposing the continuity conditions on $\psi(z)$ and $\psi'_z(z)$ at $z=\pm\frac{t}{2}$ we obtain the following set of four equations
\begin{equation}
 \left\{\begin{array}{ll}
    A_1 e^{-\kappa \frac{t}{2} }= A_2e^{-\kappa \frac{t}{2} }+B_2 e^{\kappa \frac{t}{2} },\\
     A_1 \kappa e^{-\kappa \frac{t}{2} } - \kappa \left(A_2  e^{-\kappa \frac{t}{2} }- B_2 e^{\kappa \frac{t}{2} }\right)=\frac{2}{l_0} A_1 e^{-\kappa \frac{t}{2} },\\
    A_2 e^{\kappa \frac{t}{2} }+B_2 e^{-\kappa \frac{t}{2} }  = A_3 e^{-\kappa \frac{t}{2} },\\
    \kappa \left(A_2  e^{\kappa \frac{t}{2} }- B_2 e^{-\kappa \frac{t}{2} }\right) + A_3\kappa e^{-\kappa \frac{t}{2}} = \frac{2}{l_0} A_3 e^{-\kappa \frac{t}{2} }.
   \end{array}
   \right.
   \label{set4dl}
\end{equation}
Excluding $A_1$ and $A_3$ we reduce the number of equations to two, and the eigen-value equation reads
\begin{equation}
 \frac{1}{l_0^2 \kappa^2}-\left(\frac{1}{l_0\kappa} - 1\right)^2 e^{2\kappa t }=0.
 \label{eigen2dd}
\end{equation}
Eq. (\ref{eigen2dd}) has two solutions $\kappa_{1,2}$ related to two energy eigen-values $-E_{1,2}$ as
$\kappa_{1,2}=\sqrt{2m_0 E_{1,2}/\hbar^2}$. The state `1' is even, and the state `2' is odd.
The eigen-values $-E_{1,2}$ are shown in Fig. \ref{figS2}(d) versus interlayer separation $t$.
When the interlayer separation becomes of atomic scale, $E_{1,2}$ degeneracy is lifted, doubling both the number of subbands
and the number of resonance peaks. For bilayer graphene, this results in four bands rather than two in the multilayer limit.

When an  external electric field is applied to one side of the double-layer junction, as shown in Fig. \ref{figS2}(b), the wave function reads
\begin{eqnarray}
 \psi(z)=
 \left\{\begin{array}{ll}
    A_1 e^{\kappa z}, & z<-\frac{t}{2},\\
    A_2 e^{\kappa z}+B_2 e^{-\kappa z}, & -\frac{t}{2}<z<\frac{t}{2},\\
    A_3 \mathrm{Ai}(\zeta), & z>\frac{t}{2};
   \end{array}
   \right.
\end{eqnarray}
where 
\begin{equation}
\zeta=-\frac{E_\kappa}{e {\cal E}_\mathrm{ext} z_0}+\frac{z}{z_0}, 
\label{zeta-dl}
\end{equation}
\begin{equation}
z_0=\left(\frac{\hbar^2 }{2m_0 e {\cal E}_\mathrm{ext}}\right)^{\frac{1}{3}},
\label{zeta0-dl}
\end{equation}
and $E_\kappa=-E_{1,2}$ are the new bound-state energies to be found below.
The continuity conditions imply the following equations
\begin{equation}
 \left\{\begin{array}{ll}
    A_1 e^{-\kappa \frac{t}{2} }= A_2e^{-\kappa \frac{t}{2} }+B_2 e^{\kappa \frac{t}{2} },\\
     A_1 \kappa e^{-\kappa \frac{t}{2} } - \kappa \left(A_2  e^{-\kappa \frac{t}{2} }- B_2 e^{\kappa \frac{t}{2} }\right)=\frac{2}{l_0} A_1 e^{-\kappa \frac{t}{2} },\\
    A_2 e^{\kappa \frac{t}{2} }+B_2 e^{-\kappa \frac{t}{2} }  = A_3 \mathrm{Ai}(\zeta_{+\frac{t}{2}}) ,\\
    \kappa \left(A_2  e^{\kappa \frac{t}{2} }- B_2 e^{-\kappa \frac{t}{2} }\right) + \frac{A_3}{z_0} \mathrm{Ai}'(\zeta_{+\frac{t}{2}})  = \frac{2A_3 }{l_0} 
    \mathrm{Ai}(\zeta_{+\frac{t}{2}}),
   \end{array}
   \right.
   \label{set4dlbias}
\end{equation}
where $\zeta_{+\frac{t}{2}}$ is $\zeta$ taken at $z=\frac{t}{2}$.
Note that $\mathrm{Ai}(\zeta)$ vanishes at $z\to \infty$ so that no additional boundary condition is required to keep the wave function finite.
Excluding $A_1$ and $A_3$ we reduce the number of equations to two, calculate the determinant obtaining the eigen-value equation given by
\begin{eqnarray}
\nonumber  && \frac{1}{l_0 \kappa} \left(\frac{1}{l_0 \kappa} +\frac{1}{2}+ \frac{1}{2z_0\kappa}\frac{\mathrm{Ai}'(\zeta_{+\frac{t}{2}})}{\mathrm{Ai}(\zeta_{+\frac{t}{2}})} \right)\\
\nonumber &&  -\left(\frac{1}{l_0\kappa} - 1\right)
\left(\frac{1}{l_0 \kappa} -\frac{1}{2}+ \frac{1}{2z_0\kappa}\frac{\mathrm{Ai}'(\zeta_{+\frac{t}{2}})}{\mathrm{Ai}(\zeta_{+\frac{t}{2}})} \right) e^{2\kappa t } \\
&& =0.
 \label{eigen2ddbias}
\end{eqnarray}
The eigen-values $-E_{1,2}$ are shown in Fig. \ref{figS2}(e) versus external field ${\cal E}_\mathrm{ext}$.
The applied field breaks mirror symmetry and increases the resonance splitting.
Interestingly, the field drives the lower bound state $E_1$ toward the bound state of an isolated single-delta potential, $E_0$.

Polarized double-layer can be modelled by means of a potential drop between the layers, $\Delta U$,
which can be related to the polarization field $P=\Delta U/(et)$, see Fig. \ref{figS2}(c). The wave function then reads
\begin{eqnarray}
 \psi(z)=
 \left\{\begin{array}{ll}
    A_1 e^{\kappa_+ z}, & z<-\frac{t}{2},\\
    A_2 \mathrm{Ai}(\zeta) +B_2 \mathrm{Bi}(\zeta), & -\frac{t}{2}<z<\frac{t}{2},\\
    A_3 e^{-\kappa_- z},& z>\frac{t}{2};
   \end{array}
   \right.
\end{eqnarray}
where
$\kappa_\pm=\sqrt{-2m_0\left(E_\kappa\pm\frac{\Delta U}{2}\right)/\hbar^2}$, 
and $\zeta$ is given by
\begin{equation}
\zeta=-\frac{E_\kappa t}{\Delta U z_0}+\frac{z}{z_0}, 
\label{zeta-pl}
\end{equation}
\begin{equation}
z_0=\left(\frac{\hbar^2 t}{2m_0 \Delta U}\right)^{\frac{1}{3}}.
\label{zeta0-pl}
\end{equation}
Imposing the boundary conditions we arrive at the following set of equations
\begin{equation}
 \left\{\begin{array}{ll}
    A_1 e^{-\kappa_+ \frac{t}{2} }= A_2 \mathrm{Ai}(\zeta_{-\frac{t}{2}}) +B_2 \mathrm{Bi}(\zeta_{-\frac{t}{2}}),\\
     A_1 \kappa_+ e^{-\kappa \frac{t}{2} } -\frac{1}{z_0}\left[A_2 \mathrm{Ai}'(\zeta_{-\frac{t}{2}})+ B_2 \mathrm{Bi}'(\zeta_{-\frac{t}{2}}) \right] \\
     =\frac{2}{l_0} A_1 e^{-\kappa_+ \frac{t}{2} },\\
    A_2\mathrm{Ai}(\zeta_{+\frac{t}{2}}) +B_2 \mathrm{Bi}(\zeta_{+\frac{t}{2}})  = A_3 e^{-\kappa_- \frac{t}{2} },\\
    \frac{1}{z_0}\left[A_2 \mathrm{Ai}'(\zeta_{+\frac{t}{2}})+ B_2 \mathrm{Bi}'(\zeta_{+\frac{t}{2}}) \right] + A_3 \kappa_-  e^{-\kappa_- \frac{t}{2} }\\
    = \frac{2}{l_0}A_3 e^{-\kappa_- \frac{t}{2} }.
   \end{array}
   \right.
   \label{set4dlpol}
\end{equation}
Excluding $A_1$ and $A_3$ we have a set of two equations given by
\begin{eqnarray}
 \nonumber && A_2\left[\frac{2}{l_0}\mathrm{Ai}(\zeta_{-\frac{t}{2}}) -\kappa_+ \mathrm{Ai}(\zeta_{-\frac{t}{2}}) +\frac{1}{z_0}\mathrm{Ai}'(\zeta_{-\frac{t}{2}})  \right]\\
 \nonumber && + B_2 \left[\frac{2}{l_0}\mathrm{Bi}(\zeta_{-\frac{t}{2}}) -\kappa_+ \mathrm{Bi}(\zeta_{-\frac{t}{2}}) +\frac{1}{z_0}\mathrm{Bi}'(\zeta_{-\frac{t}{2}})  \right]=0\\
 \nonumber && A_2\left[\frac{2}{l_0}\mathrm{Ai}(\zeta_{+\frac{t}{2}}) -\kappa_- \mathrm{Ai}(\zeta_{+\frac{t}{2}}) - \frac{1}{z_0}\mathrm{Ai}'(\zeta_{+\frac{t}{2}})  \right]\\
 \nonumber && + B_2 \left[\frac{2}{l_0}\mathrm{Bi}(\zeta_{+\frac{t}{2}}) -\kappa_- \mathrm{Bi}(\zeta_{+\frac{t}{2}}) -\frac{1}{z_0}\mathrm{Bi}'(\zeta_{+\frac{t}{2}})  \right]=0.\\
\end{eqnarray}
The eigen-value equation then reads
\begin{eqnarray}
 \nonumber && \left[\frac{2}{l_0}\mathrm{Ai}(\zeta_{-\frac{t}{2}}) -\kappa_+ \mathrm{Ai}(\zeta_{-\frac{t}{2}}) +\frac{1}{z_0}\mathrm{Ai}'(\zeta_{-\frac{t}{2}})  \right]\\
 \nonumber && \times\left[\frac{2}{l_0}\mathrm{Bi}(\zeta_{+\frac{t}{2}}) -\kappa_- \mathrm{Bi}(\zeta_{+\frac{t}{2}}) -\frac{1}{z_0}\mathrm{Bi}'(\zeta_{+\frac{t}{2}})  \right] \\
 \nonumber && - \left[\frac{2}{l_0}\mathrm{Bi}(\zeta_{-\frac{t}{2}}) -\kappa_+ \mathrm{Bi}(\zeta_{-\frac{t}{2}}) +\frac{1}{z_0}\mathrm{Bi}'(\zeta_{-\frac{t}{2}})  \right] \\
 \nonumber && \times \left[\frac{2}{l_0}\mathrm{Ai}(\zeta_{+\frac{t}{2}}) -\kappa_- \mathrm{Ai}(\zeta_{+\frac{t}{2}}) - \frac{1}{z_0}\mathrm{Ai}'(\zeta_{+\frac{t}{2}})  \right] \\ && =0.
 \label{eigenpol}
\end{eqnarray}
Eq. (\ref{eigenpol}) can be solved with respect $E_\kappa$ to find two solutions $-E_{1,2}$ shown in Fig. \ref{figS2}(f).
The energy splitting $E_1-E_2$ increases with increasing polarization, as expected.

\section{S2. Field emission from a bare metal}

\begin{figure*}
 \includegraphics[width=0.6\textwidth]{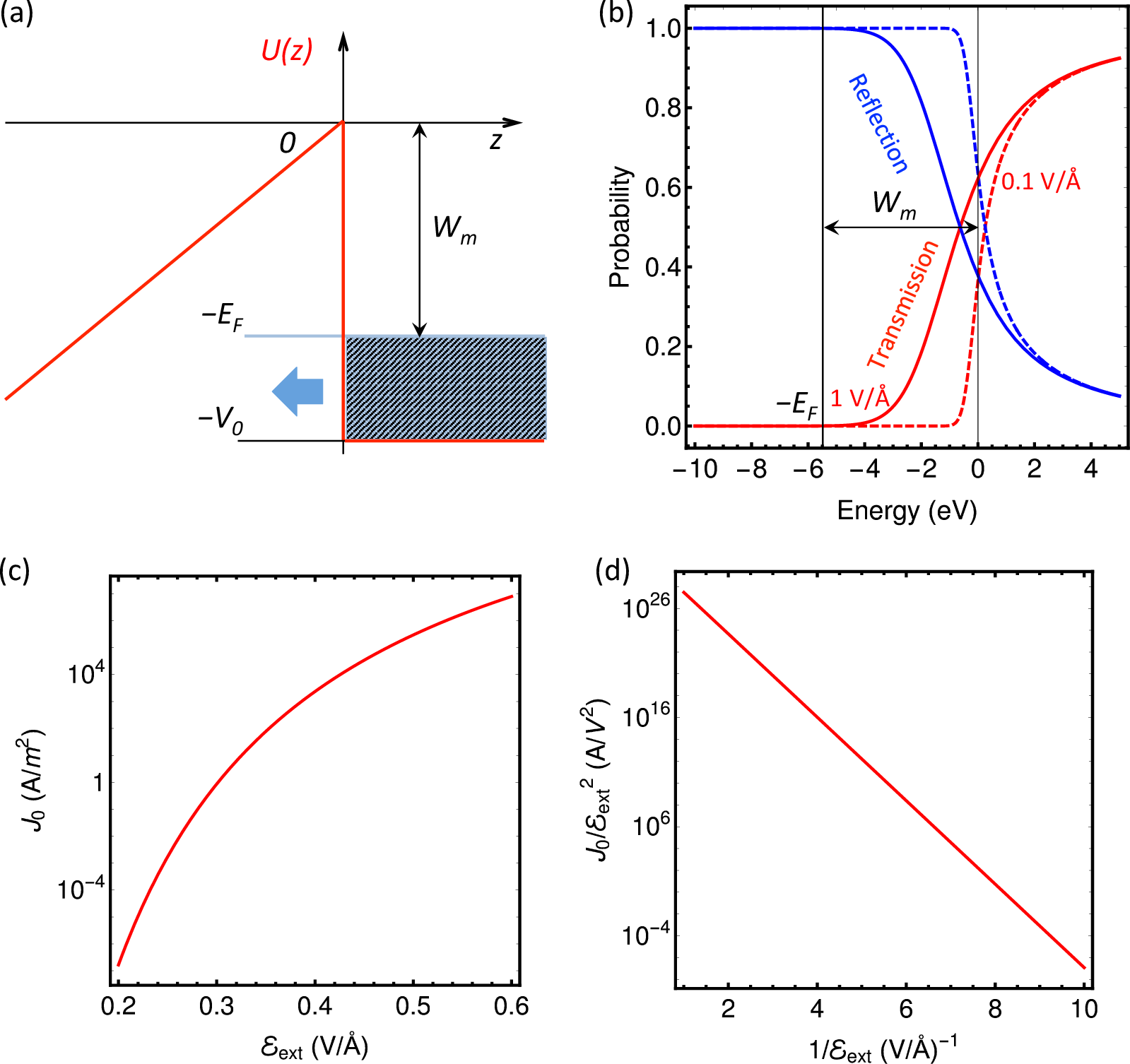}
 \vspace{1cm}
 \caption{Electron field emission from a metal without graphene. (a) Potential profile $U(z)$ in an external field ${\cal E}_\mathrm{ext}$.
 Here $W_m=E_F$ is the metal work function, $E_F>0$ is the Fermi energy (measured from the vacuum level), $V_0>0$ is the conduction-band depth,
 and the energy interval between $-V_0$ and $-E_F$ is occupied by electrons.
 (b) Transmission and reflection probabilities versus energy referenced to from the vacuum level.
  Solid curves: ${\cal E}_\mathrm{ext}=1$ V/\AA; dashed curves: ${\cal E}_\mathrm{ext}=0.1$ V/\AA.
 (c) Field emission current density as a function of ${\cal E}_\mathrm{ext}$.
 Electron-electron interactions  (e.g., image-charge effects) are neglected, which overestimates the field required to obtain a measurable current.
 (d) Fowler-Nordheim representation showing a linear relation between   $\ln(J_0/{\cal E}_\mathrm{ext}^2)$ and $1/{\cal E}_\mathrm{ext}$.
 \label{figS3}
}
\end{figure*}

Field electron emission from a bare metal is usually calculated within the Wentzel-Kramers-Brillouin (WKB) approximation.
However, we use an exact solution here, because resonant tunneling occurring in the presence of graphene is not tractable semiclassically.
The potential profile is shown in Fig. \ref{figS3}(a).
The wave function can be written in terms of the Airy functions $\mathrm{Ai}$ and $\mathrm{Bi}$ as
\begin{eqnarray}
 \psi(z)=
 \left\{\begin{array}{ll}
    A_1\left[\mathrm{Ai}(\zeta)+i\mathrm{Bi}(\zeta)\right], & z<0,\\
    A_3 e^{-ikz} +B_3 e^{ikz},& z>0;
   \end{array}
   \right.
\end{eqnarray}
where $\zeta$ is given by 
\begin{equation}
\zeta=-\frac{E_k}{e{\cal E}_\mathrm{ext} z_0}+\frac{z}{z_0}, 
\label{zetaE}
\end{equation}
\begin{equation}
z_0=\left(\frac{\hbar^2}{2m_0 e{\cal E}_\mathrm{ext}}\right)^{\frac{1}{3}}.
\label{zeta0E}
\end{equation}

 The total electron energy reads
\begin{equation}
 E_\mathrm{tot}=E_k+\frac{\hbar^2 k_\parallel^2}{2m},
\end{equation}
where $\hbar k_\parallel$ is the in-plane momentum, and
\begin{equation}
 E_k=-V_0+\frac{\hbar^2 k^2 }{2m},
\end{equation}
where $\hbar k$ is the out-of-plane momentum.
The amplitudes $A_1$ and $A_3$ can be found by matching the wave function and its derivative at $z=0$ as
\begin{eqnarray}
&& A_1\left[\mathrm{Ai}(\zeta_0)+i\mathrm{Bi}(\zeta_0)\right]=A_3 +B_3\\
&& \frac{A_1}{z_0}\left[\mathrm{Ai}'(\zeta_0)+i\mathrm{Bi}'(\zeta_0)\right]=-ik A_3  +ik B_3,
\end{eqnarray}
where $\zeta_0$ is $\zeta$ taken at $z=0$, and $\mathrm{Ai}'(\zeta)$, $\mathrm{Bi}'(\zeta)$ are the derivatives of the Airy functions.
The incident and reflected probability current densities respectively read
\begin{equation}
 j_i=\frac{\hbar k}{m}\lvert B_3 \rvert^2,\quad j_r=\frac{\hbar k}{m}\lvert A_3 \rvert^2,
\end{equation}
whereas the transmitted probability current density is given by
\begin{eqnarray}
 j_t & = & \frac{\hbar}{mz_0} \lvert A_1 \rvert^2\left[\mathrm{Ai}(\zeta)\mathrm{Bi}'(\zeta)-\mathrm{Ai}'(\zeta)\mathrm{Bi}(\zeta) \right]\\
 & =& \frac{\hbar}{\pi mz_0} \lvert A_1 \rvert^2.
\end{eqnarray}
The transmission and reflection probabilities can be written as
\begin{equation}
 T= \frac{j_t}{j_i},\quad R= \frac{j_r}{j_i}.
\end{equation}
By convention, we take $k<0$ for the incident current propagating toward $z\to -\infty$, and we set $B_3=1$ for normalization.
The probabilities are shown in Fig. \ref{figS3}(b) for two different ${\cal E}_\mathrm{ext}$.

\begin{figure*}
 \includegraphics[width=0.9\textwidth]{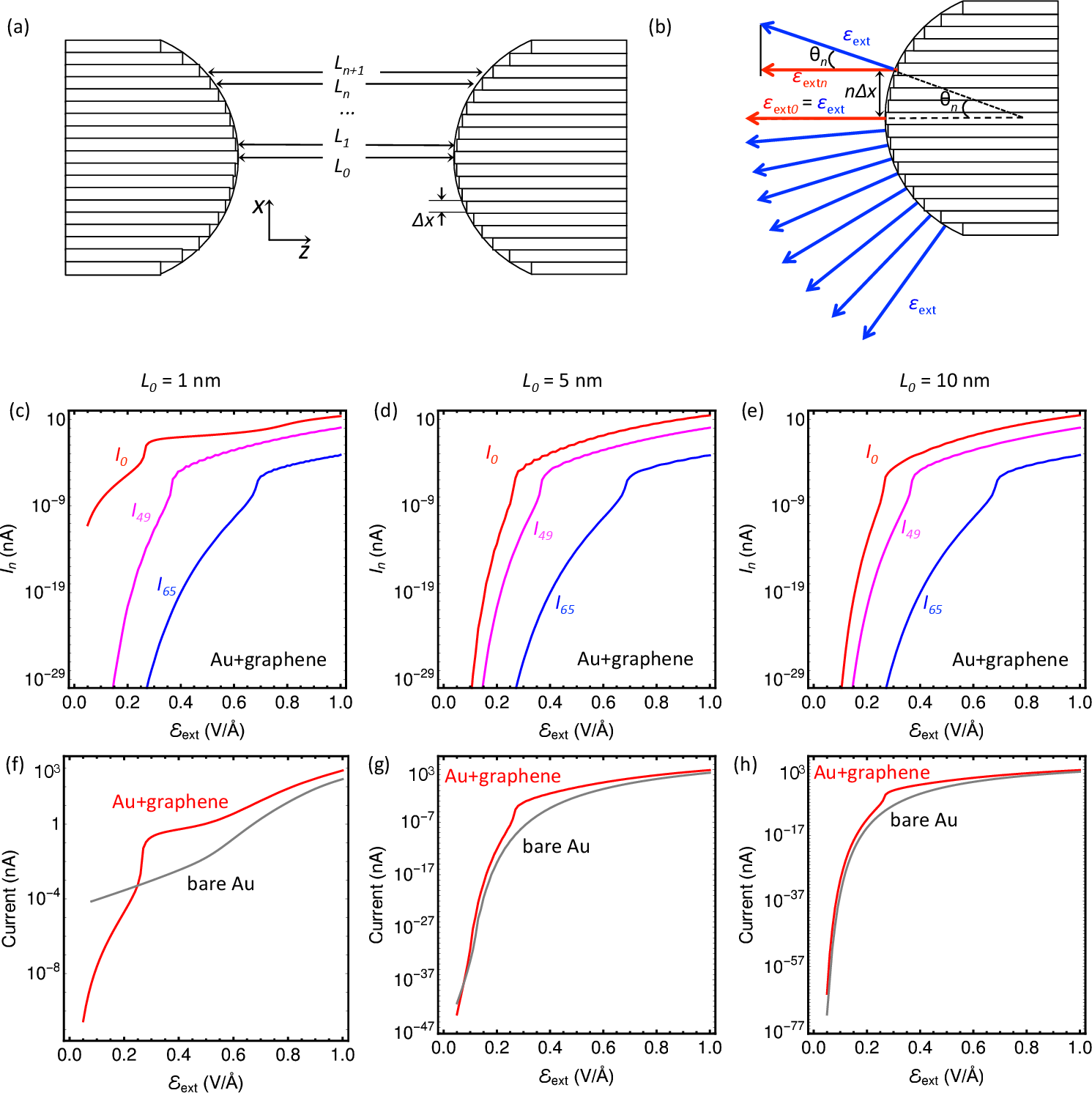}
 \caption{Electrode surface curvature. (a) The electrodes are modeled as two nanorods with identical semicircular tips of radius
 $r_0=10$ nm and height of $\Delta y= 10$ nm. Each tip is partitioned into elementary emitters of area $\Delta x \Delta y$,
 separated by different path lengths $L_n$, see Eq. (\ref{Ln}) with $\Delta x=1.42$\AA, and $n$ is the integer 0,...,70. 
 (b) The true electric field ${\cal E}_\mathrm{ext}$ (blue arrows) is everywhere normal to the metallic surface.
 For the partitioned surface, the effective field driving emission from the $n$-th emitter is the component along its local normal,
 ${\cal E}_\mathrm{ext} \cos\theta_n$.
 (c,d,e) Emission current $I_n$ for three representative elementary emitters and three electrode separations, see Eq. (\ref{In}).
 The central channel $I_0$ always dominates, indicating strongest emission from the apex. Oscillations in $I_n$ arise from backscattering between the electrodes.
 (f,g,h) Total current obtained by summing all elementary contributions. Gray curves show the corresponding current for bare gold electrodes.
 The transition between direct tunneling and field emission is evident in (f) near $\sim 0.5$ V/\AA\,  and in (g) near $\sim 0.1$ V/\AA\, 
 as the intersection of gray and red curves. For $L_0>10$ nm, direct tunneling is negligible, so red and gray curves coincide in both low- and high-field limits,
 while a clear difference persists at the resonance field.
 \label{figS4}
}
\end{figure*}

The total electrical current density is calculated by integrating over occupied states in the metal weighted by the probability current density as
\begin{eqnarray}
 J_0 & = & 2e\int \frac{dk}{2\pi} \int \frac{d^2k_\parallel}{(2\pi)^2} v_k T(k)\\
 & = & 2e\int \frac{dk}{2\pi} \int \frac{d^2k_\parallel}{(2\pi)^2} j_t(k).
\end{eqnarray}
where $v_k=\hbar k/m$. The subscript ``0'' in $J_0$ denotes the absence of graphene.
The integration limits can be found from the following conditions
\begin{eqnarray}
 E_\mathrm{tot} & \leq  & -E_F \\
k_\parallel & \leq & \sqrt{\frac{2m}{\hbar^2}\left(-E_k-E_F\right)},
\end{eqnarray}
and
\begin{eqnarray}
-E_k-E_F & \geq & 0 \\
k & \leq & \sqrt{\frac{2m}{\hbar^2}\left(V_0-E_F\right)},
\end{eqnarray}
see Fig. \ref{figS3}(a) for $V_0$ and $E_F$.

Switching to the cylindrical coordinates and changing the integration variable we have
\begin{eqnarray}
 \nonumber J_0 & =& \frac{e}{(2\pi)^2}\int\limits_{-V_0}^{-E_F}dE_k \frac{2m}{\hbar^2}\frac{j_t\left(\sqrt{\frac{2m}{\hbar^2}\left(V_0+E_k\right)}\right)}{\sqrt{\frac{2m}{\hbar^2}\left(V_0+E_k\right)}}\\
 & &\times   \int\limits_0^{\sqrt{\frac{2m}{\hbar^2}\left(-E_F-E_k\right)}} dk_\parallel k_\parallel \\
 \nonumber & = & \frac{e}{8\pi^2}\left(\frac{2m}{\hbar^2}\right)^{\frac{3}{2}}\int\limits_{-V_0}^{-E_F} dE_k \frac{-E_F-E_k}{\sqrt{V_0+E_k}}\\
 && \times j_t\left(\sqrt{\frac{2m}{\hbar^2}\left(V_0+E_k\right)}\right).
 \label{j0}
\end{eqnarray}
The field emission current density described by equation (\ref{j0}) follows the well known dependence on external field, as shown in Fig. \ref{figS3} (c,d).
We use $J_0$ given by equation (\ref{j0}) to normalize the field emission current in the presence of graphene.

\section{S3. Finite-size device simulations}

In the field emission regime, transmission is typically modeled through a triangular barrier, assuming the anode is effectively at infinity.
In this limit, the transmission probability is governed by the barrier shape rather than the cathode-anode separation.
However, when the device is small or the field is weak, the separation becomes critical, distinguishing field emission from direct tunneling.
Here, we consider the length explicitly including $L$ into the solution given by
\begin{eqnarray}
\label{wfL}
 \psi(z)=
 \left\{\begin{array}{ll}
    A_0 e^{-ipz} +B_0 e^{ipz},& z<-L; \\
    A_1\mathrm{Ai}(\zeta_1)+B_1 \mathrm{Bi}(\zeta_1), & -L<z<0,\\
    A_2 \mathrm{Ai}(\zeta_2)+B_2 \mathrm{Bi}(\zeta_2), & 0<z<d,\\
    A_3 e^{-ikz} +B_3 e^{ikz},& z>d;
   \end{array}
   \right.
\end{eqnarray}
\begin{eqnarray}
&& \zeta_1 =  -\frac{E_k}{e{\cal E}_\mathrm{ext}z_1}+\frac{z}{z_1},\quad z_1=\left(\frac{\hbar^2}{2me{\cal E}_\mathrm{ext}}\right)^{\frac{1}{3}},\\
&& \zeta_2 =  -\frac{E_k d}{\Delta V z_2}+\frac{z}{z_2},\quad z_2=\left(\frac{\hbar^2 d}{2m\Delta V}\right)^{\frac{1}{3}},\\
&& k=\sqrt{\frac{2m}{\hbar^2}\left(V_0 + E_k\right)},\\
&& p=\sqrt{\frac{2m}{\hbar^2}\left(V_0+\Delta V + e{\cal E}_\mathrm{ext}L + E_k\right)}.
\end{eqnarray}
By convention, we take $A_0=0$ and set $B_3=1$ for normalization.
The remaining coefficients are found by matching the wave function and its derivative at $z=-L$, $z=0$, and $z=d$ as
\begin{eqnarray}
&& \label{se1} B_0 e^{-ipL}= A_1\mathrm{Ai}(\zeta_{1L})+B_1 \mathrm{Bi}(\zeta_{1L}), \\
&& \label{se2} ipB_0 e^{-ipL} = \frac{1}{z_1}\left[ A_1\mathrm{Ai}'(\zeta_{1L})+B_1 \mathrm{Bi}'(\zeta_{1L})\right], \\ 
&& \label{se3} A_1\mathrm{Ai}(\zeta_{10})+B_1 \mathrm{Bi}(\zeta_{10})  = A_2\mathrm{Ai}(\zeta_{20})+B_2 \mathrm{Bi}(\zeta_{20}),\\
\nonumber &&  \frac{1}{z_1}\left[A_1 \mathrm{Ai}'(\zeta_{10})+B_1\mathrm{Bi}'(\zeta_{10})\right] \\
\nonumber && - \frac{1}{z_2}\left[A_2 \mathrm{Ai}'(\zeta_{20})+B_2 \mathrm{Bi}'(\zeta_{20})\right]\\
&&\label{se4} =\frac{2}{l_0}\left[ A_2\mathrm{Ai}(\zeta_{20})+B_2 \mathrm{Bi}(\zeta_{20}) \right],\\
&&\label{se5}  A_2\mathrm{Ai}(\zeta_{2d})+B_2 \mathrm{Bi}(\zeta_{2d})  = A_3 e^{-ikd} +B_3 e^{ikd},\\
&& \nonumber   \frac{1}{z_2}\left[A_2 \mathrm{Ai}'(\zeta_{2d})+B_2 \mathrm{Bi}'(\zeta_{2d})\right]\\
&& \label{se6} =  -ik A_3 e^{-ikd} +ik B_3 e^{ikd},
\end{eqnarray}
where $\zeta_{1L}=\zeta_1(z=L)$, $\zeta_{10}=\zeta_1(z=0)$, $\zeta_{1d}=\zeta_1(z=d)$, $\zeta_{20}=\zeta_2(z=0)$, $\zeta_{2d}=\zeta_2(z=d)$.
Solving Eqs. (\ref{se1}--\ref{se6}) yields $A_{1,2,3}$ and $B_{0,2}$. 
The incident, reflected, and transmitted probability current densities are, respectively, given by
\begin{equation}
 j_i=\frac{\hbar k}{m}\lvert B_3 \rvert^2,\quad j_r=\frac{\hbar k}{m}\lvert A_3 \rvert^2, \quad j_t  = \frac{\hbar p}{m}\lvert B_0 \rvert^2.
\end{equation}

\begin{figure*}
 \includegraphics[width=0.9\textwidth]{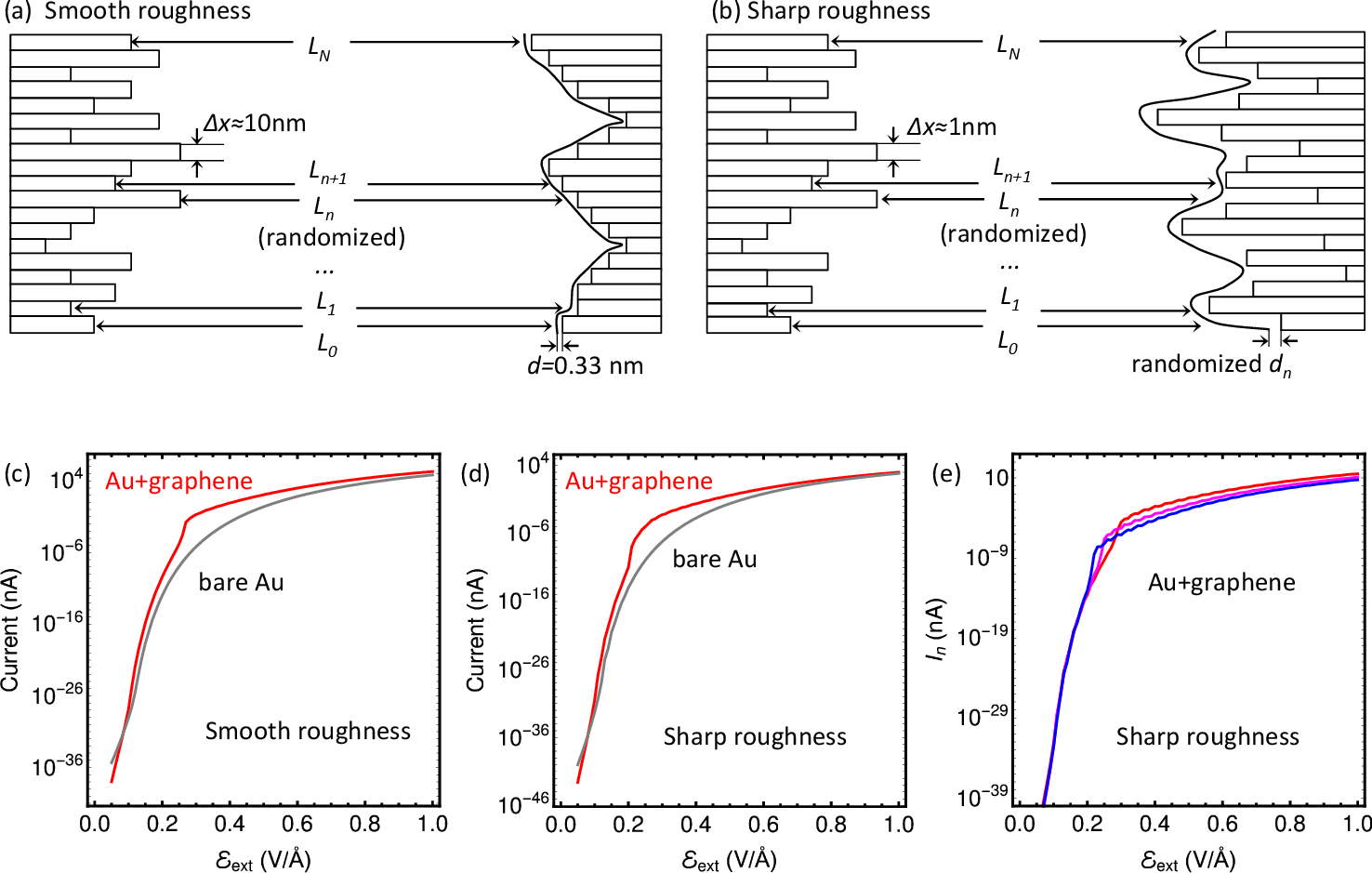}
  \vspace{1.2cm}
 \caption{Electrode surface roughness. (a) Smooth roughness, where the graphene layer conforms to the underlying metal surface profile
  and the graphene-metal separation remains approximately uniform.  (b) Sharp roughness, where the graphene-metal separation is randomized.
 (c) Total emission current for smooth roughness, showing the same resonance feature as in the perfectly flat case. 
 Gray curve corresponds to bare gold electrodes.
 Configuration: 100 channels, $\Delta x\Delta y= 100$ nm$^2$ with $L$ randomized in the interval $4.5$--$5.5$ nm.
 (d) For sharp roughness, the resonance is broadened but remains visible. Configuration: 
 100 channels, $\Delta x\Delta y= 1$ nm$^2$; $L$ and $d$ are independently randomized over  $4.9$--$5.1$ nm
 and $0.23$--$0.43$ nm, respectively.
 (e) Emission currents $I_n$ for three representative channels contributing to (d). 
 Random resonance shifts in each channel flatten the overall resonance feature in (d).
 \label{figS5}
}
\end{figure*}

If graphene is removed, then Eqs. (\ref{se1}--\ref{se6}) are reduced to the following set
\begin{eqnarray}
&& \zeta_1 =  -\frac{E_k}{e{\cal E}_\mathrm{ext}z_1}+\frac{z}{z_1},\quad z_1=\left(\frac{\hbar^2}{2me{\cal E}_\mathrm{ext}}\right)^{\frac{1}{3}},\\
&& \zeta_2 =  -\frac{E_k d}{\Delta V z_2}+\frac{z}{z_2},\quad z_2=\left(\frac{\hbar^2 d}{2m\Delta V}\right)^{\frac{1}{3}},\\
&& k=\sqrt{\frac{2m}{\hbar^2}\left(V_0 + E_k\right)},\\
&& p=\sqrt{\frac{2m}{\hbar^2}\left(V_0+\Delta V + e{\cal E}_\mathrm{ext}L + E_k\right)}.
\end{eqnarray}
By convention, we take $A_0=0$ and set $B_3=1$ for normalization.
The remaining coefficients are found by matching the wave function and its derivative at $z=-L$, $z=0$, and $z=d$ as
\begin{eqnarray}
&& \label{e01} B_0 e^{-ipL}= A_1\mathrm{Ai}(\zeta_{1L})+B_1 \mathrm{Bi}(\zeta_{1L}), \\
&& \label{e02} ipB_0 e^{-ipL} = \frac{1}{z_1}\left[ A_1\mathrm{Ai}'(\zeta_{1L})+B_1 \mathrm{Bi}'(\zeta_{1L})\right], \\ 
&& \label{e03} A_1 \mathrm{Ai}(\zeta_0)+B_1 \mathrm{Bi}(\zeta_0) =A_3 +B_3\\
&& \label{e04} \frac{1}{z_0}\left[A_1 \mathrm{Ai}'(\zeta_0)+B_1 \mathrm{Bi}'(\zeta_0)\right]=-ik A_3  +ik B_3,
\end{eqnarray}
where $\zeta_0$ is $\zeta(z=0)$ determined by Eq. (\ref{zetaE}).
The incident, reflected, and transmitted probability current densities are defined in the same way as before.
The finite-size solution allows us to model curved and uneven anode and cathode surfaces.

Let us start from smoothly curved electrodes, Fig. \ref{figS4}.
By partitioning each surface into small, finite elementary emitters, we obtain a set of channels, each with a well-defined length $L_n$. 
If the radius of curvature ($r_0$) is constant and identical for both electrodes, then $L_n$ is given by
\begin{equation}
 L_n= L_0 + 2r_0-2\sqrt{r_0^2-\left(n\Delta x\right)^2},
 \label{Ln}
\end{equation}
where $L_0$ is the length of the central (shortest) channel, and $\Delta x$ is the channel width.
The typical curvature of the gold nanorods considered here is $r_0=10$ nm.
The natural width of an elementary emitter is set by graphene's lattice scale, we take $\Delta x = 1.42$\AA.

In addition to the length change, we have to model the field texture which is not collinear in the case of curved surface.
The z-component of the electric field which drives the current between the electrodes 
is different for each elementary emitter and can be written as
\begin{eqnarray}
{\cal E}_{\mathrm{ext}n} & = & {\cal E}_\mathrm{ext} \cos\theta_n\\
& = & {\cal E}_\mathrm{ext}\sqrt{1-\left(\frac{n\Delta x}{r_0}\right)^2}.
 \label{eEn}
\end{eqnarray}

Using the results of the previous section, the emission current for the $n$-th channel can be written as
\begin{eqnarray}
 \nonumber I_n & = & \frac{e\Delta x\Delta y}{8\pi^2}\left(\frac{2m}{\hbar^2}\right)^{\frac{3}{2}}\int\limits_{-V_0}^{-E_F} dE_k \frac{-E_F-E_k}{\sqrt{V_0+E_k}}\\
 && \times j_t\left(\sqrt{\frac{2m}{\hbar^2}\left(V_0+E_k\right)},L_n,{\cal E}_{\mathrm{ext}n}\right),
 \label{In}
\end{eqnarray}
where $\Delta x\Delta y$ is the elementary cross section of a channel.
The curvature of the field lines is not included in this approximation.
Note that $V_0$ is defined differently depending on whether a graphene layer is present.
The total current is then the sum over all channels, $I=\sum_n I_n$.
The channel at the apex of a electrode (i.e. $I_n$ with $n=0$) dominates in field-emission because (i) the driving electric field is maximized
(${\cal E}_{\mathrm{ext}0}={\cal E}_{\mathrm{ext}}$) and (ii) the channel length in minimized.

Finally, we address rough emitting surfaces. Two scenarios are relevant (Fig. \ref{figS5}):
(1) smooth roughness, where the graphene conforms to the underlying metal topography and the graphene-metal separation $d$ remains essentially constant;
and (2) sharp roughness, where graphene can sag over local troughs, randomizing $d$ within a finite range. Unlike the curved-tip case,
in the planar geometry the driving electric field is the same for all elementary emitters.
Hence, the emission current for the $n$-th channel reads
\begin{eqnarray}
 \nonumber I_n & = & \frac{e\Delta x\Delta y}{8\pi^2}\left(\frac{2m}{\hbar^2}\right)^{\frac{3}{2}}\int\limits_{-V_0}^{-E_F} dE_k \frac{-E_F-E_k}{\sqrt{V_0+E_k}}\\
 && \times j_t\left(\sqrt{\frac{2m}{\hbar^2}\left(V_0+E_k\right)},L_n,d_n\right),
 \label{Ind}
\end{eqnarray}
where $L_n$ and $d_n$ are randomized within the respective intervals $L_\mathrm{max}<L_n<L_\mathrm{min}$, $d_\mathrm{max}<d_n<d_\mathrm{min}$.

\end{document}